\documentclass[12pt]{iopart}

\makeatletter
\@namedef{ver@amsmath.sty}{}
\makeatother
\usepackage{amstext}
\usepackage{graphicx, booktabs}

\usepackage{siunitx}
\pdfmapfile{+sansmathaccent.map}

\begin{document}

\title{A practical guide to Terahertz imaging using thermal atomic vapour.}

\author{Lucy A. Downes$^1$, Lara Torralbo-Campo$^2$, Kevin J. Weatherill$^1$}

\address{$^1$ Department of Physics, Durham University, South Road, Durham. DH1 3LE, UK}
\address{$^2$ Physikalisches Institute, Eberhard Karls Universität Tübingen, Auf der Morgenstelle 14, 72076 Tübingen, Germany}

\ead{
k.j.weatherill@durham.ac.uk}
\vspace{10pt}
\begin{indented}
\item[\today]
\end{indented}

\begin{abstract}
This tutorial aims to provide details on the underlying principles and methodologies of atom-based terahertz imaging techniques. Terahertz imaging is a growing field of research which can provide complementary information to techniques using other regions of the electromagnetic spectrum. Unlike infrared, visible and ultraviolet radiation, terahertz passes through many everyday materials, such as plastics, cloth and card. Compared with images formed using lower frequencies, terahertz images have superior spatial resolution due to the shorter wavelength, while compared to x-rays and gamma rays, terahertz radiation is non-ionising and safe to use. The tutorial begins with the basic principles of terahertz to optical conversion in alkali atoms before discussing how to construct a model to predict the fluorescent spectra of the atoms, on which the imaging method depends. We discuss the practical aspects of constructing an imaging system, including the subsystem specifications. We then review the typical characteristics of the imaging system including spatial resolution, sensitivity and bandwidth. We conclude with a brief discussion of some potential applications.  
\end{abstract}

\section{Introduction.}

\subsection{Using Atoms as Sensors}

Atoms in dilute vapour can make very effective sensors \cite{kitching2011,Degen2017}. Atoms have no moving parts to wear out, they are relatively unperturbed by inter-atomic interactions, their energy levels are sensitive to applied fields, and crucially, each atom of the same isotope is identical. This final point means that atomic sensors are in effect pre-calibrated and measurements made using them are reproducible and should be, at least in principle, traceable to the SI system of measurements. Atomic systems already provide a platform for precision clocks~\cite{Knappe2004,Ludlow2015}, gyroscopes~\cite{kornack2005,Fang2012}, magnetometers~\cite{Budker2007,Schwindt04}, gravimeters~\cite{Janvier2022,Narducci2022} and gradiometers~\cite{Perry:20}. Many atom-based sensors achieve optimal performance by using laser-cooling techniques to create very cold atomic samples~\cite{vovrosh2022}, and although efforts are ongoing to simplify and miniaturise such apparatus~\cite{McGilligan2017,McGilligan2020}, laser cooling inevitably introduces significant experimental complexity and cost to the setup. In contrast, for atomic sensors where inhomogeneous Doppler broadening is not a problem, setups using room-temperature thermal vapour contained within glass cells provide a simpler route to applications.

In recent years, a new class of atomic sensors has emerged based on Rydberg atoms~\cite{Adams2020}. Rydberg atoms are atoms from which one or more electrons has been promoted to a highly excited state~\cite{Rydberg_Physics} and are characterised by their large size, electric polarizability and strong electric dipole transitions between neighboring states. Rydberg atoms are therefore very sensitive to both static and oscillating electric fields. Furthermore, using the technique of electromagnetically induced transparency (EIT)~\cite{Boller1991,Finkelstein2022} it is possible to achieve coherent optical detection of Rydberg atoms in a thermal vapour~\cite{Mohapatra2008} and therefore have a direct readout of the effects of applied electric fields on the atoms. Using Rydberg EIT it has been possible to detect microwave electric fields with unprecedented sensitivity~\cite{Tanasittikosol_2011,Sedlacek2012,Holloway2014,superhet}. This successful technique has subsequently been applied across a huge frequency range spanning from the ultra low frequency~\cite{Yuan2020} to the terahertz range~\cite{Chen:22}.

\subsection{Imaging with Atoms}

In principle any measurement, or set of measurements, which allows the spatial distribution of a field to be determined can be considered an `imaging' technique. In that sense, any of the above-mentioned sensors can be used for imaging if repeated measurements can be made at different spatial positions over a time period where the field of interest does not change. For example, measurements using Rydberg atoms can achieve sub-wavelength imaging of microwave electric fields by moving, or rastering, the sensor with respect to the measured field~\cite{Holloway2014b}. Furthermore, atomic vapours can also be used in imaging modalities where the full field  is captured in a single shot. For example, room temperature alkali vapour can be used to measure the magnetic component of microwave fields by detecting the population transfer between atomic hyperfine ground states~\cite{Bohi2012}. Using this technique the field above microwave circuits has been imaged in a single shot at frame rates of around 10~Hz and spatial resolution below 100 micrometers~\cite{Horsley2015}.  
Some other atom-based imaging techniques employ standard optical methods in combination with atomic-vapour line filters to provide narrow-band transmission and strong out-of-band rejection for imaging at specific wavelengths~\cite{Logue:22} thereby allowing the signal of interest to be extracted from unwanted background light. Atomic line filters have been used for imaging applications ranging from astronomical solar monitoring~\cite{samnet} and imaging of jet plumes~\cite{Kudenov:20} to microscopy for single DNA molecule detection~\cite{Uhland2015}.

In this tutorial review we concentrate on the imaging of terahertz (THz) fields using Rydberg atoms in room-temperature vapour \cite{Downes2020}. Similar to atomic line filters, terahertz imaging using atomic vapour provides a signal at a specific frequency corresponding to an atomic transition, whilst offering strong out-of-band rejection of other frequencies. However, unlike atomic filters, this method up-converts the frequency of interest, in this case from the terahertz range, to another, higher optical frequency. This burgeoning technique has already enabled unprecedented speed and sensitivity for full-frame terahertz imaging where more traditional terahertz imaging technologies struggle for performance.

\subsection{The Terahertz Gap}

The terahertz range of the electromagnetic spectrum is typically defined as the frequency range between 0.1 and 10~THz, or the wavelengths from 3~mm to $30\,\si{\micro\metre}$. This range is often referred to as the `THz gap' because it falls between the ranges of electronic, microwave devices and semiconductor-based infrared devices. The terahertz range has historically been one of the least explored regions of the electromagnetic spectrum due to the relative lack of convenient sources and detectors. Fortunately, the last two decades have seen a tremendous development of new terahertz technology \cite{Davies_2002, Tonouchi2007,Lewis2014} to overcome this situation. 

The motivation to develop terahertz technologies is, in large part, to enable the exploitation of the unique properties of terahertz radiation~\cite{Dhillon2017,Bruendermann2012}. Terahertz radiation combines the ability to deeply penetrate many common materials such as plastics, cloth and paper whilst also being non-ionising, thereby providing a safe alternative to x-rays for some applications. Furthermore, terahertz radiation provides a useful tool for materials identification, as the absorption bands of many technologically-useful materials and molecules fall within the terahertz range. Terahertz radiation is strongly absorbed by polar liquids such as water, making it useful for passive hydration monitoring~\cite{Gente2015}. However, the same property means that terahertz radiation is strongly attenuated in the Earth's atmosphere because of the water vapour content~\cite{Coutaz2018} and is therefore unsuitable for long-distance free-space communication technology. Nevertheless terahertz-frequency fields are proposed for local data transmission within 10 metres \cite{Ishigaki2012} and for wireless communications at high altitude and in space (aircraft to satellite, or satellite to satellite) \cite{Shian2013}. The properties of terahertz radiation enable a multitude of other applications ranging from security screening \cite{Tonouchi2007}, and non-destructive evaluation for quality control in industry \cite{Federici:20,Brinkmann2017,Naftaly2019}, drugs and explosives detection \cite{Davies2008,Neumaier2014,Liu2016},  biology and medicine \cite{Siegel2004,Yang2016,Amini2021} to ultrafast spectroscopy \cite{Lucia2003}, astrophysics and atmospheric science \cite{Ritchter2015,Rezac2015}. Many of these applications involve terahertz imaging as they require the spatial structure of the target material to be recorded~\cite{Richter2010,Hagelschuer:16}.

Terahertz imaging techniques can be crudely divided into two types: schemes involving single-pixel detection and those using focal plane array detectors. For most single-pixel imaging schemes, one needs to translate the target through the imaging plane to build up an image of the object of interest. In general, this is a very time-consuming process and image collection can take many hours. However, the spatial resolution can be very good (sub-diffraction limited) as it can be defined by the size of a small aperture translated through the imaging system. However, better resolution comes at the cost of longer acquisition times. Single pixel imaging systems can use many different combinations of terahertz source and detector; we provide a brief review of terahertz sources in Section~\ref{sec:thzsources} but on detectors we simply note that they typically rely on a heating effect and therefore cryogenic detectors are faster and more sensitive than those operating at room temperature. By far the most successful and popular method for terahertz imaging is time-domain spectroscopy (TDS), which uses a mode-locked titanium:sapphire laser system and GaAs antennae to produce, and then detect, temporally short and spectrally broadband terahertz pulses~\cite{TDS_tutorial}. TDS provides both amplitude and phase information of the terahertz field, as well as spectral information, and provides the underlying technology behind commercial imaging systems from Teraview~\cite{teraview} and Menlo Systems~\cite{menlo-tds}. 
Another single-pixel terahertz-imaging method uses self-mixing in quantum cascade lasers (QCLs)~\cite{Dean_2014}. Here, the same QCL is used as both the source and the detector and back-reflections of the terahertz field from the imaging target are fed back into the QCL facet and detected electrically. This method is particularly suited to stand-off imaging over extended distances, but like other single-pixel schemes, requires time-consuming raster scanning. 

Focal plane array detectors are comprised of multiple sensors and, unlike single pixel detectors, they can capture a full-field image in a single shot. The charged coupled device (CCD) camera is a classic example of a focal place array operating in the optical and near infrared range. CCD cameras cannot operate in the terahertz range, however arrays of detectors such as microbolometers~\cite{Lee:05,Lee:06}, field-effect transistors (FETs)~\cite{Qin2017} or carbon nanotubes~\cite{Suzuki2016} can be used instead. Such arrays can operate up to video frame rates of around 50~Hz; however, their sensitivities are limited, and scaling to large pixel numbers can be challenging. Nonetheless, commercial systems based on uncooled microbolometers are available from companies such as INO~\cite{INO} and i2s~\cite{i2s}. Despite significant progress in recent years, true real-time, full-field operation remains an unachieved goal for many terahertz imaging applications~\cite{Mittleman18}, therefore, novel atom-based terahertz imaging offers a viable alternative to the above-mentioned technologies as it provides the fastest frame rates currently possible. Table~\ref{tab:thz-imagers} summarises the pros and cons of various terahertz imaging systems. Thermal atomic vapours offer a good combination of properties i.e. very high speed, full-field imaging at room temperature but unlike TDS systems only operate at a single frequency.

\begin{table}
\begin{center}
\begin{tabular}{c}
\includegraphics[width=11.5cm]{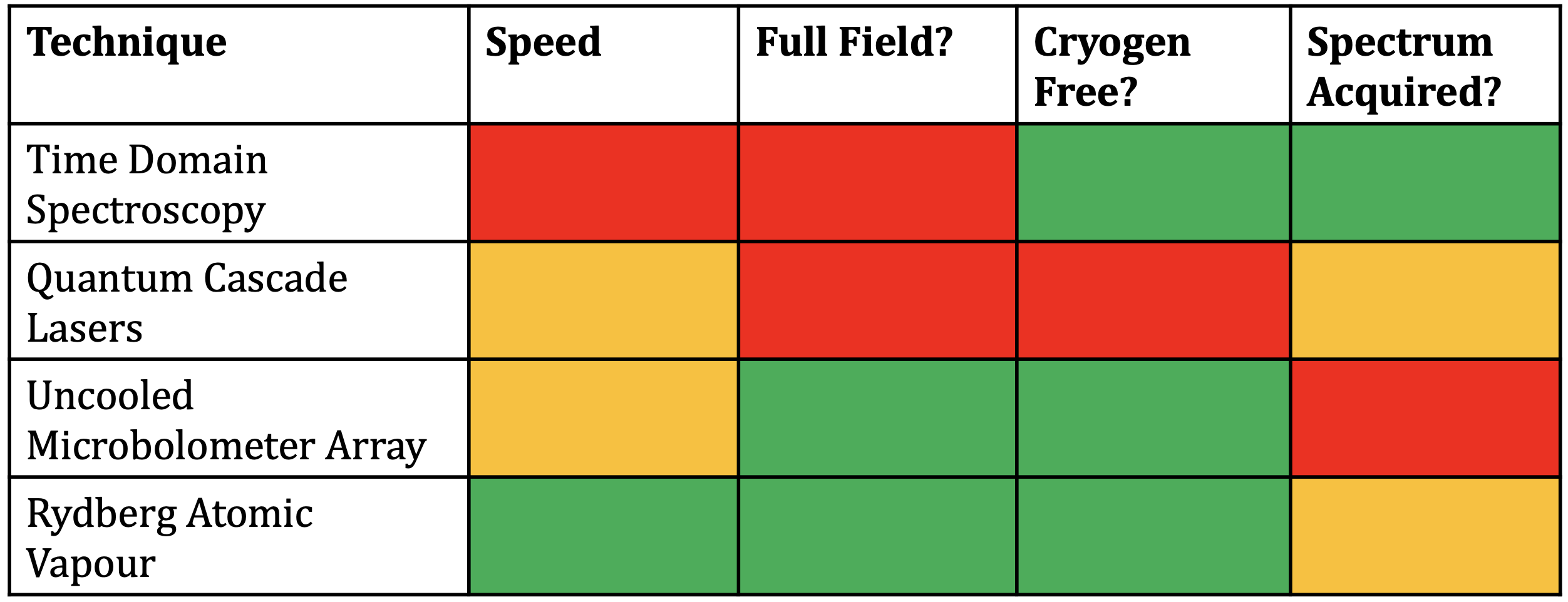} 
\end{tabular}
\caption{\label{tab:thz-imagers} Summary of the pros and cons of different terahertz imaging techniques. Red colour signifies poor performance, orange means medium performance and green signifies good performance for a particular attribute. Atom-based terahertz imagers have a good range of attributes.}
\end{center}
\end{table}

\subsection{Layout of the article}

This article is structured as follows. In Section~\ref{sec:theory}, we present the basic concept of atom-based terahertz imaging and discuss the atomic structure and atom-light interactions relevant for the terahertz to optical conversion process at the heart of the technique. We then discuss the level structure and excitation schemes within two commonly used alkali atoms, rubidium and caesium, in Section~\ref{sec:atomic_structure}. We outline a simple model to predict the fluorescence spectra from the excited atoms in Section~\ref{sec:model} and using the information from these model spectra, we can begin to select efficient levels for imaging, as described in Section~\ref{sec:state_choice}. 
The experimental setup is described in detail in Section~\ref{sec:exptsetup}. We cover the key individual components such as terahertz sources (Section~\ref{sec:thzsources}), terahertz optics (Section~\ref{sec:thzoptics}), atomic vapour cells (Section~\ref{sec:cells}), excitation lasers (Section~\ref{sec:lasers}) and cameras (Section~\ref{sec:cameras}). We explain how the elements are put together and how the experiment is controlled in Section~\ref{sec:expt_control}. Section~\ref{sec:results} discusses the results and characterisation of an example of this atom-based imaging system. In Section~\ref{sec:resolution} we discuss how the spatial resolution can be determined and the limitations on this. In Section~\ref{sec:temporal} we assess the temporal resolution of the system and discuss how this is typically limited by the optical camera. In Section~\ref{sec:sensitivity} we discuss measurements of the sensitivity of the system and the minimum detectable intensity and in Section~\ref{bandwidth} we discuss the spectral bandwidth of the system. Finally, in Section~\ref{sec:applications}, we discuss some potential applications of this high speed terahertz imaging system and outline future opportunities and challenges for the technology. 

\section{Terahertz to optical conversion and how to pick a `good' imaging transition}\label{sec:theory}

\subsection{Basic concept of atom-based terahertz imaging}

\begin{figure}[hbt]
\begin{centering}
\includegraphics[width=0.9\textwidth]{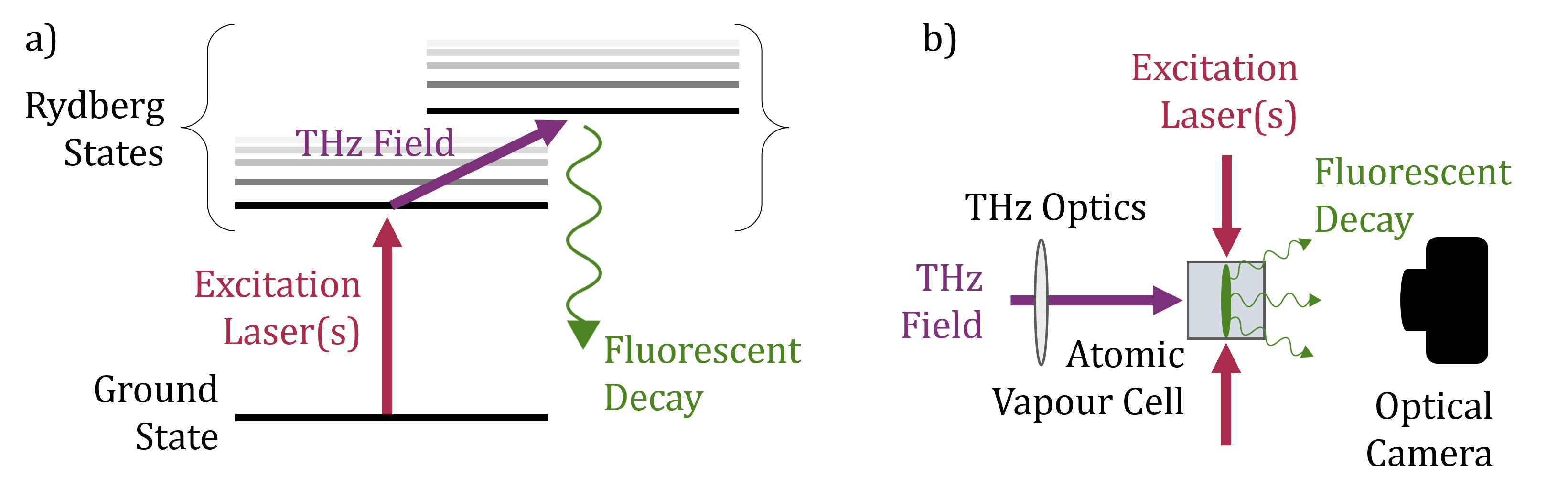}
\caption{\textbf{Introduction to atom-based THz imaging.} a) Energy level diagram of generic atom. The atom is excited to a Rydberg state using lasers (red arrow). Once in the Rydberg state, a terahertz field (purple arrow) can transfer atomic population to an adjacent Rydberg state from which it can spontaneously decay, emitting optical fluorescence (green arrow). Ideally, the fluorescent decay from the terahertz-coupled state will have a different wavelength to the initial Rydberg state, signalling the presence of the terahertz field. b) Basic experimental setup. Atoms within the vapour cell are excited into a light sheet using lasers. A terahertz field is applied from an orthogonal direction, then the resulting fluorescence is collected and imaged using an optical camera.}
\label{fig:basic_idea}
\end{centering}
\end{figure}

Terahertz imaging using Rydberg atoms in room temperature atomic vapour \cite{Wade2017,Downes2020} is a relatively new technique which converts difficult to detect terahertz photons into easy to detect optical photons. This type of `upconversion' allows terahertz field distributions to be mapped onto visible light fields, which can then be collected using an optical camera. Figure~\ref{fig:basic_idea} demonstrates the basic principles of the technique: a) shows a generic atomic energy level scheme, where excitation lasers are used to resonantly promote population from the atomic ground state to an excited state with large principal quantum number ($n>10$) known as a `Rydberg' state. An incoming terahertz field can then resonantly transfer population to another neighbouring Rydberg state. Population can decay spontaneously from both Rydberg states, emitting optical fluorescence. However, if appropriate care is taken in the choice of states, as discussed later in Section~\ref{sec:state_choice}, the optical fluorescence spectrum emitted in the presence of the terahertz field will be very different from the fluorescence spectrum seen without an applied terahertz field. Therefore the terahertz-induced fluorescence can be isolated from the background, providing an image of the incident field. More detailed energy level schemes for the laser excitation of the two main atomic species used, caesium and rubidium, are shown later in Section~\ref{sec:atomic_structure}. 
Figure~\ref{fig:basic_idea}b) shows a basic schematic of how this technique works in practice. Atoms contained within an atomic vapour cell are excited using lasers, creating a thin sheet of excited Rydberg atoms within the cell. A terahertz field is applied from an orthogonal direction to the lasers causing the excited atoms within the vapour to fluoresce at a different wavelength and this fluorescence can be isolated and captured by an optical camera. The spatial pattern of the optical fluorescence from the atomic vapour matches the spatial pattern of the applied terahertz field, meaning that the optical image collected at the camera effectively provides an image of the terahertz field. 

Although, in principle, terahertz imaging using Rydberg atoms can work for any atomic transition in the appropriate frequency range, it is certainly not true that all transitions, and all atomic species, will work equally well. There are many factors that influence the choice of states and laser excitation schemes. This section explains some of the factors affecting the quality of terahertz imaging and how we can make informed choices when planning experiments to maximise the performance of the imaging technique. 

\subsection{Relevant atomic structure}\label{sec:atomic_structure}

We will not cover basic atomic structure here, referring the reader instead to other sources~\cite{Foot:1080846,Rydberg_Physics}. Instead, we begin by considering the frequency difference, $\Delta \nu$, between adjacent atomic states, 1 and 2, in a monovalent atom.
\begin{equation}
    \displaystyle{\Delta \nu = \left|c R_{\infty}\left(\frac{1}{{n_2^*}^2} - \frac{1}{{n_1^*}^2} \right)\right|, }
\end{equation}
where $c$ is the speed of light, $R_{\infty}$ is the Rydberg constant, $n_m^* = n_m - \delta_m$, $n_m$ and $\delta_m$ are the principal quantum number and quantum defect of state $m$ respectively.
For states with low orbital angular momentum ($\ell < 3$) and typical values of $\delta_m$, we find that $\Delta \nu \approx 1$~THz when $n_1, n_2 \approx 20$. This is why we require highly-excited `Rydberg' states to access transitions in the terahertz range.

In theory any atomic species could be used to perform terahertz imaging using the framework laid out here, but we will limit our considerations to rubidium (Rb) and caesium (Cs). Rb and Cs are commonly used in many atomic physics experiments because the laser wavelengths required to address atomic transitions from the ground state are readily available. Both Rb and Cs also have high vapour pressure at room temperature meaning that experiments can be performed in a vapour cell with little to no heating required.
Throughout this tutorial, atomic energy states will be referred to using the notation $n\ell_{j}$ where $n$ is the principal quantum number, $\ell$ is the orbital angular momentum quantum number denoted using S, P, D, F for $\ell = 0,1,2,3$ etc. and $j = |\ell\pm1/2|$ is the total angular momentum quantum number. 

\begin{figure}[hbt]
\begin{centering}
\includegraphics[width=\textwidth]{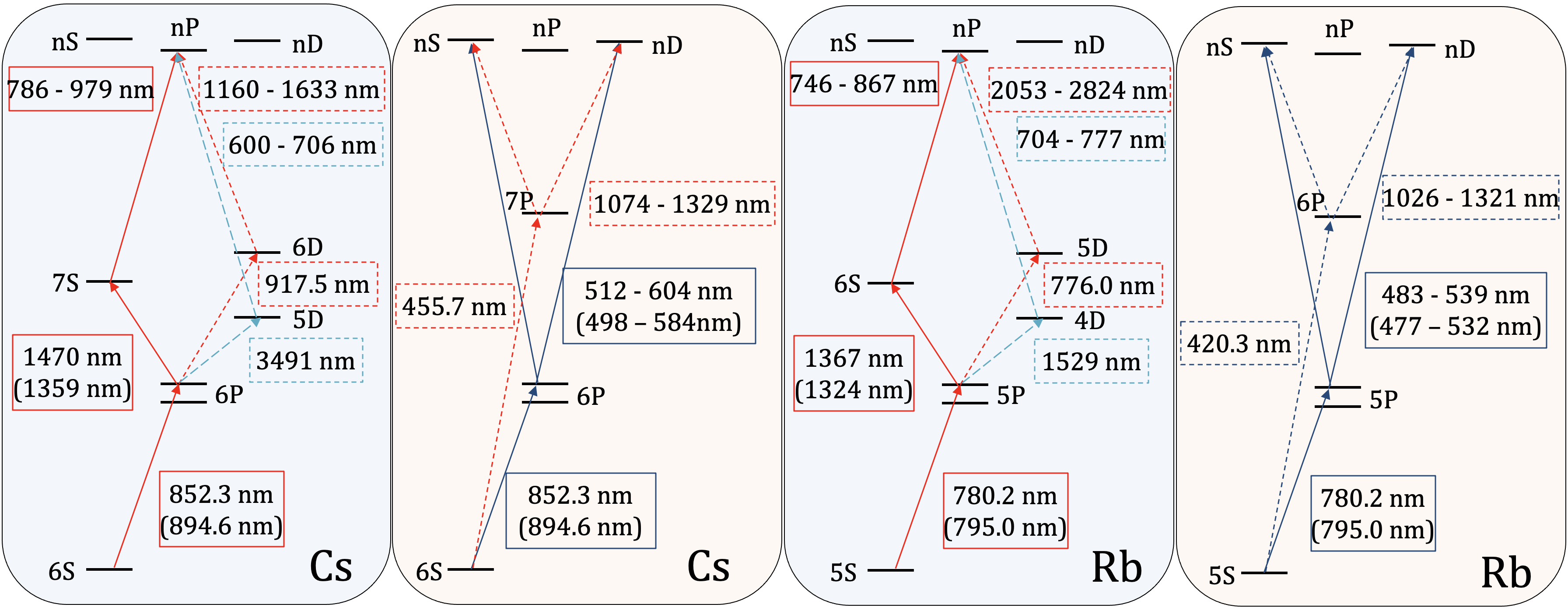}
\caption{\textbf{Laser excitation schemes in Cs and Rb.} Some of the most convenient laser excitation pathways to Rydberg states. Three photon schemes are shown in pale blue boxes and two-photon schemes in beige boxes. The wavelengths quoted assume transitions to and from the highest $j$ value except for those in parentheses which are transitions to/from a P$_{1/2}$ state. The wavelength ranges of the final step transitions represent excitation to states in the range $n=10-30$. The laser technology used to generate these wavelengths is discussed further in Table~\ref{table:lasers}.}
\label{fig:schemes}
\end{centering}
\end{figure}

There are very many excitation schemes possible, however for simplicity we present only schemes where there are appropriate laser sources readily available. For example, it is possible to excite Rydberg states directly from the ground state \cite{wang2017}, but this requires high power ultraviolet lasers which are expensive and relatively difficult to work with. More commonly, multi-photon excitation schemes are used which reduce the power and photon energy required for any single laser. Examples of the wavelengths and powers used in these multi-photon schemes are discussed in greater detail in Section~\ref{sec:lasers}. Figure~\ref{fig:schemes} shows some examples of two- and three-photon excitation schemes in both Cs (left) and Rb (right). The quoted vacuum wavelengths assume transitions to and from the highest $j$ value of the states except for those in parentheses which are transitions to/from a P$_{1/2}$ state. The quoted wavelength ranges of the final step transitions represent excitation to states with principal quantum number in the range $n=10-30$ as terahertz transitions are typically accessible from states with principal quantum numbers in this range.

The excitation scheme used determines the angular momentum quantum number $\ell$ of the initial Rydberg state. Dipole allowed transitions, such as those induced by an electromagnetic field, can only occur between states with $\Delta \ell = 1$, so an S state ($\ell=0$) can only be coupled to a P state ($\ell=1$). 
This means that the excitation scheme used determines the value of $\ell$ of the states used for imaging. 
For example a two-photon excitation scheme (beige boxes, Figure~\ref{fig:schemes}) would see population transferred from the S ground state to an intermediate P state, from which the second laser could be used to couple to either an S or a D state. The terahertz transition could then couple to either a P or an F state. 
Using a three-photon scheme (blue boxes, Fig.~\ref{fig:schemes}) gives potentially more flexibility and greater choice of $\ell$ states. Ultimately, regardless of which set of states and associated laser excitation scheme is used, the goal is to allow the efficient transfer of atomic population from the ground state up to a Rydberg state, such that the ensemble becomes sensitive to the application of a terahertz field.

\subsection{Simple Model for Predicting Fluorescent Spectra}\label{sec:model}

For this terahertz imaging scheme to work effectively we need to select a pair of atomic states to act as the initial Rydberg and terahertz coupled state which satisfy certain criteria:
\begin{itemize}
    \item Frequency separation equal to the terahertz frequency of interest.
    \item Large transition dipole matrix element (strong transition).
    \item Large probability for the emission of optical fluorescence.
    \item Good contrast between the optical emission of the two states.
\end{itemize}
The first two requirements are easily investigated using open-source software such as the Alkali Rydberg Calculator (ARC)~\cite{Sibalic2017}. ARC is a Python program which calculates the properties of alkali Rydberg atoms and allows us to search for atomic transitions that satisfy these criteria. However, finding states that satisfy the latter two criterion requires further work as they cannot be calculated analytically.
In order to predict the fluorescent spectra of an ensemble of excited Rydberg atoms we need to take into account the stochastic and probabilistic nature of spontaneous decay. As we will see, we also need to include blackbody-radiation-induced transitions and collisional population transfer to achieve good agreement between predicted and experimentally observed spectra. 

\begin{figure}[ht]
\begin{centering}
\includegraphics[width=0.95\textwidth]{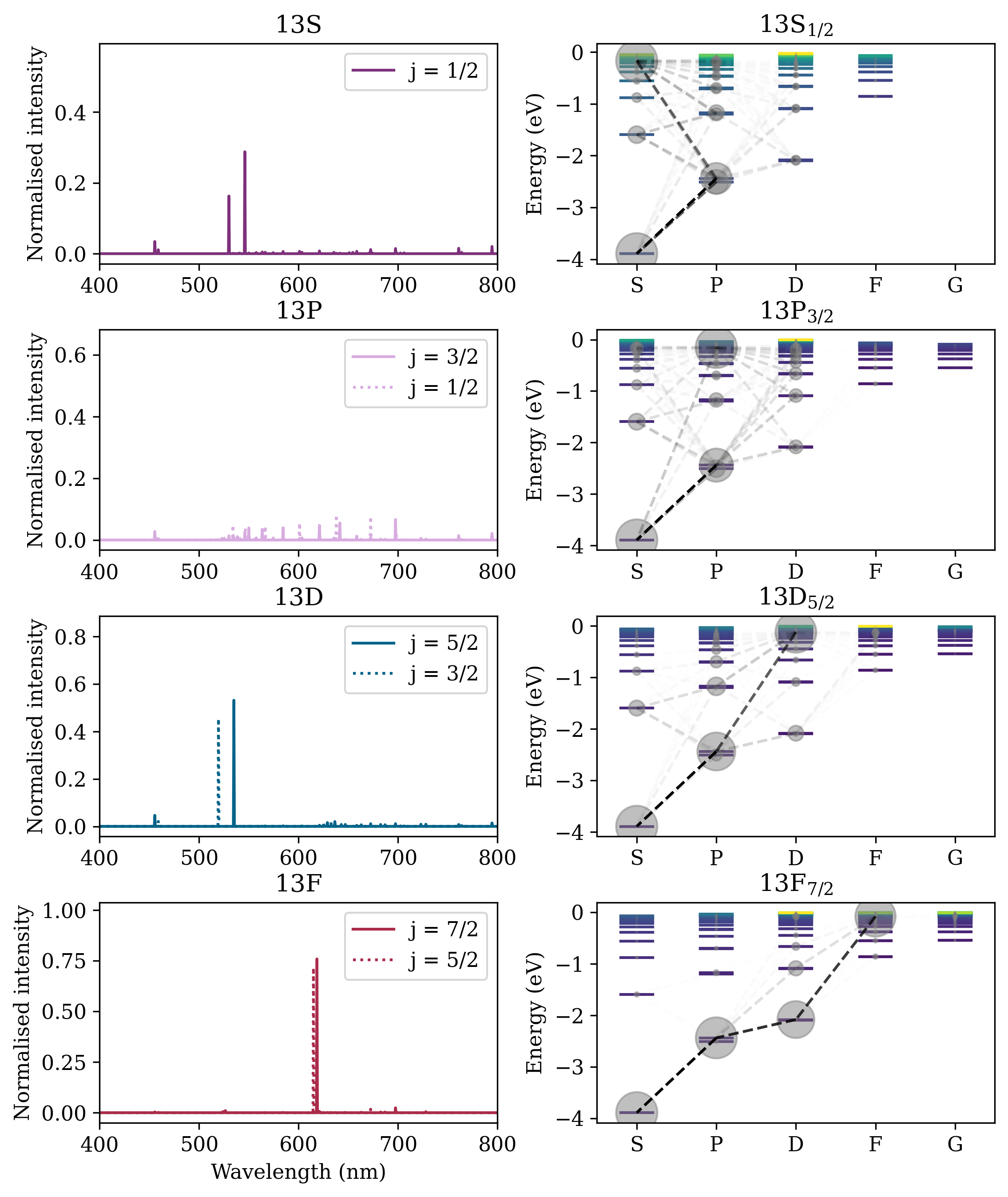}
\caption{\textbf{Results of Monte Carlo fluorescence model} Predicted fluorescence spectra (left) and decay diagrams showing the decay pathways taken by atoms in the Monte Carlo model (right) for decay from a Rydberg state ($n=13$) in Cs at 300~K. The rows show results for varying values of $l$ ($\ell = 0$ to 3, top to bottom). In the modelled spectra, results for both hyperfine states are shown where applicable. In the decay pathway diagrams only the upper hyperfine states are shown for simplicity. The opacity of the dashed lines represents the likelihood of the pathway occurring in the model. The size of the circles represents the amount of population that passes through the corresponding state.}
\label{fig:decay_paths}
\end{centering}
\end{figure}

To this end, we construct a Monte Carlo simulation of fluorescent decay from any given Rydberg level. We first create a `look-up' table of the transition probabilities from each atomic state to every other atomic state within a given range ($n<80,\,\ell<6$). 
We then use this look-up table in combination with random number generation to `choose' which transitions an excited atom undergoes until it reaches the ground state. For every transition we record the wavelength of any photons emitted, and repeat many times (usually $>10,000$) to build up a histogram of emitted photons. This gives a prediction of the spectral characteristics of the fluorescence from the vapour when excited to any given Rydberg state. 
We can also use this model to see which `path' an atom took to reach the ground state and hence see which decay pathways are causing strong emission lines. The results of using this simulation to model decay out of the $n=13, \ell \leq 3$ states in Cs are shown in Fig.~\ref{fig:decay_paths}.

\begin{figure}[hbt]
\begin{centering}
\includegraphics[width=0.9\textwidth]{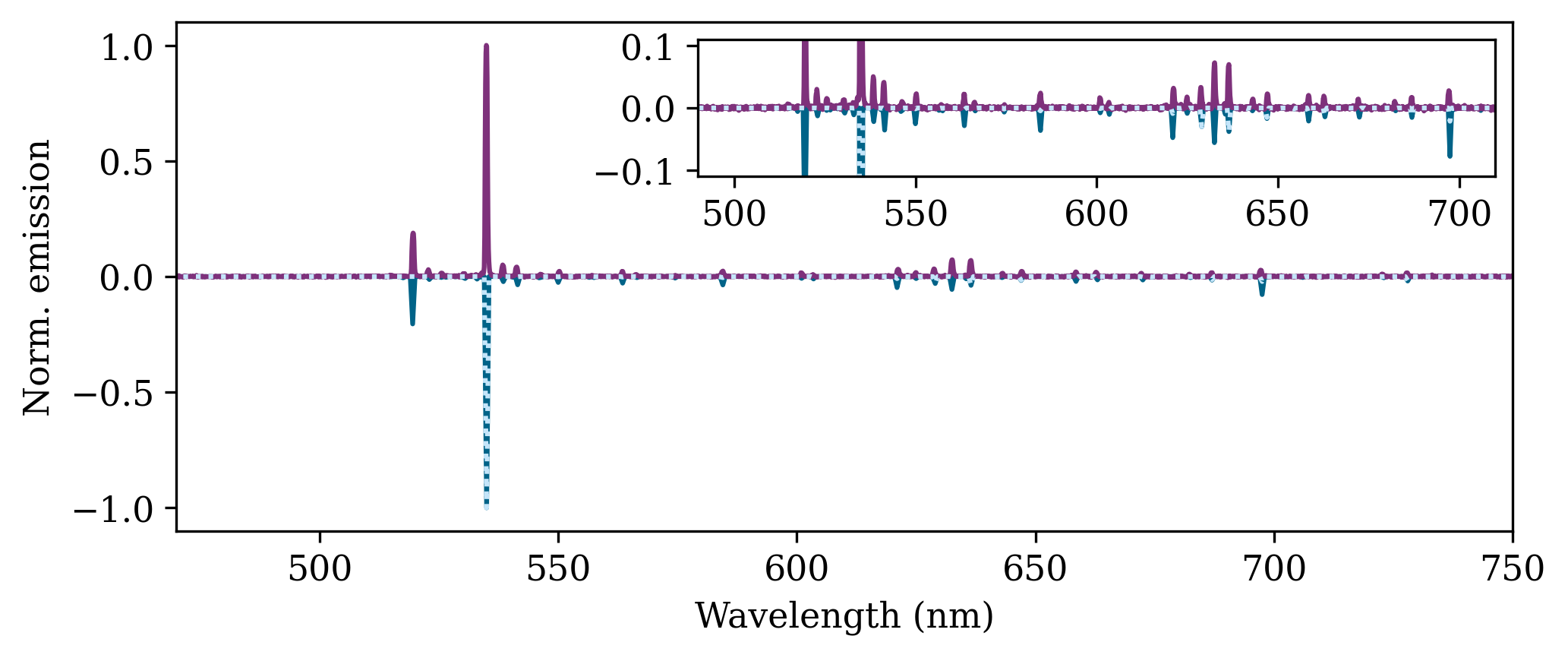}
\caption{\textbf{Comparison of Monte-Carlo simulation with data} Measured fluorescence spectrum (purple line) for the $13\rm{D}_{5/2}$ state in Cs, taken with an OceanOptics Flame USB spectrometer. The data shows good agreement with the simulated spectrum (dark blue line) in which population remaining in the  $14\rm{P}_{1/2}$ state and collisional transfer to the $13\rm{D}_{3/2}$ state is accounted for. The predicted spectrum for $13\rm{D}_{5/2}$ alone (light blue dotted line) shows that there are many features of the measured spectrum not captured by this, motivating the need to consider atomic population decaying from elsewhere.}
\label{fig:spectrum}
\end{centering}
\end{figure}
To check the validity of this model and its ability to inform our choice of `good' states for THz imaging we compare the result of the simulation to the spectrally resolved fluorescence from a Rydberg state ($13\mathrm{D}_{5/2}$ in Cs), shown in Fig.~\ref{fig:spectrum}. The spectrum is measured using an OceanOptics Flame USB spectrometer. 
This shows that although the model does accurately predict the wavelength of the strongest emission line (in this case at $535\,\si{\nano\metre}$) to get a better match to the data we need to consider that not all population will be transferred from the initial $14\mathrm{P}_{3/2}$ state by the THz field and that some $j$-mixing collisional transfers occur within the vapour, most notably to the lower hyperfine state ($13\mathrm{D}_{5/2}\rightarrow 13\rm{D}_{3/2}$). This effect is added manually into the model as a phenomenological effect.

\subsection{Selecting the Optimum Transitions}\label{sec:state_choice}

For the purposes of a terahertz imaging experiment the parameter of interest is the contrast between the fluorescence emitted by two Rydberg states coupled by a terahertz field. 
Ideally we want to be able to use an optical filter to select fluorescence that is predominantly due to decay from the terahertz-coupled Rydberg state (`THz on') and remove unwanted background fluorescence from the initial Rydberg state (`THz off'). 
We also want to have a relatively high probability of emitting a photon within this wavelength range of interest (a signal photon) to give a good terahertz-to-optical conversion efficiency. The simple Monte Carlo model described in the previous section enables us to extract both of these important parameters without experimentally exploring the many hundreds of possible terahertz transitions available.
The eventual measured fluorescence signal $f_{\rm{on}} - f_{\rm{off}}$ will be proportional to the number of atoms in the `THz on' state $n_{\rm{on}}$ multiplied by the difference in probabilities $p_{\rm{on}} - p_{\rm{off}}$ of emitting a signal photon from the two states. Hence we choose the $p_{\rm{on}} - p_{\rm{off}}$ quantity from the model as our figure of merit (FoM) when deciding on which states will be suitable for imaging.
The dependence on $n_{\rm{on}}$ indicates that we also need to consider the amount of population that can be transferred by the terahertz field. 
We expect $n_{\rm{on}} \propto \Omega_{\rm{THz}}^2$ where $\Omega_{\rm{THz}}$ is the Rabi frequency of the terahertz field. Since power, $P\propto \Omega^2$, 
$n_{\rm{on}}$ (and therefore $f_{\rm{on}} - f_{\rm{off}}$) will depend on the square of the dipole matrix element of the terahertz transition $\mathbf{d}_{\rm{off}\rightarrow\rm{on}}^2$ and linearly on the power of the terahertz field $P_{\rm{THz}}$.
This relationship allows us to use our fluorescence model along with knowledge of the dipole matrix elements to predict the signal strength of any given transition and hence choose the best pair of states for terahertz imaging in a given frequency range. This can be combined with knowledge about the terahertz source output power to see whether a certain transition is a viable choice.

\begin{figure}
\centering
\includegraphics[width = 0.95\linewidth]{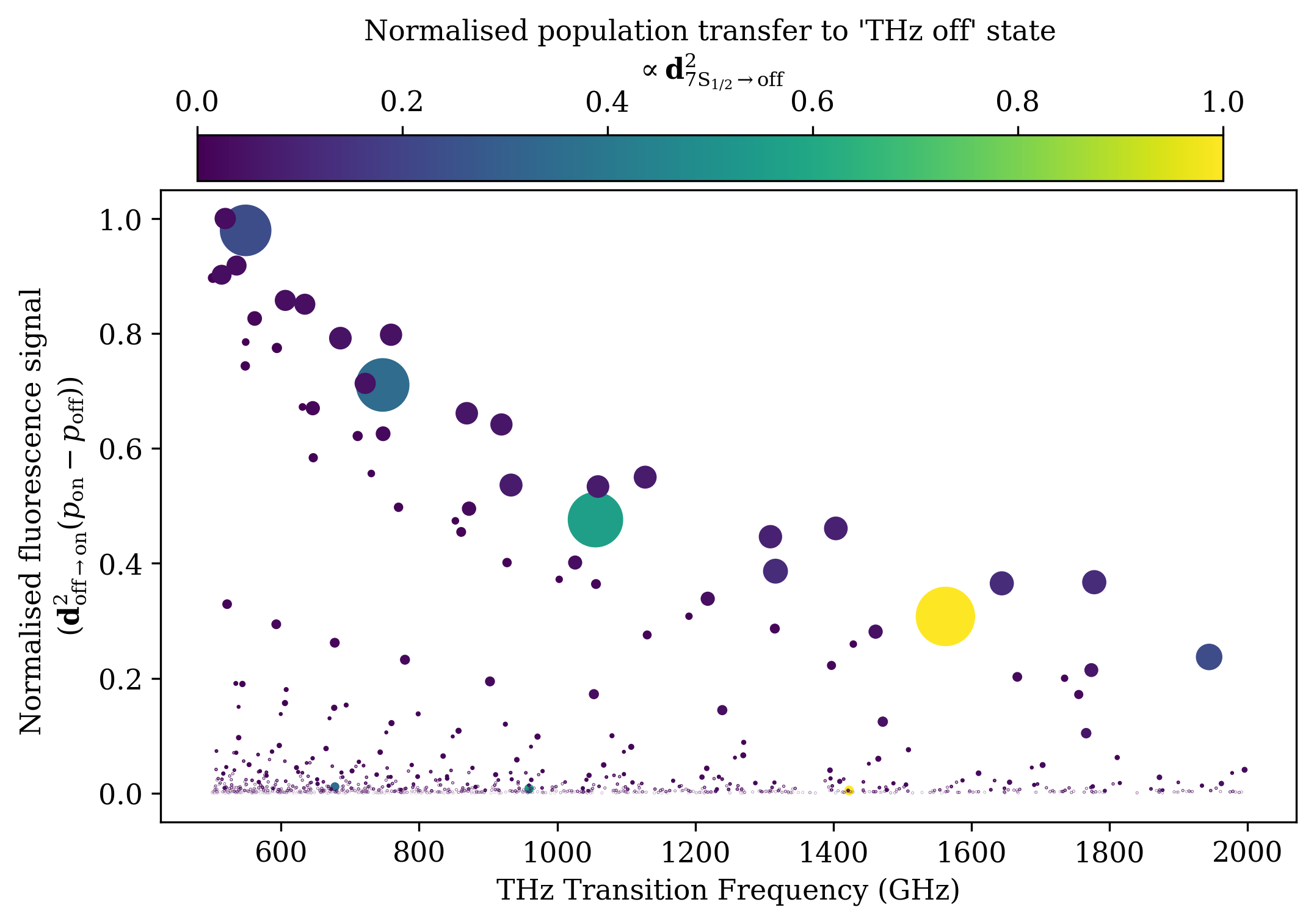}
\caption{\textbf{Evaluation of THz transitions for imaging.} Using the Monte-Carlo model we can predict the figure-of-merit (FoM) for all transitions in a given THz range and compare them. Here we consider the 3-photon excitation scheme via the $7\rm{S}_{1/2}$ state in Cs. The scatter points represent the frequency of the transition against its predicted fluorescence signal strength. 
The colour of the points is related to the square of the dipole matrix element of the transition to the initial Rydberg state ($\mathbf{d}_{\rm{7S}_{1/2}\rightarrow \rm{off}}^2$), giving an idea of how much population can be transferred to the `THz off' state for a fixed laser power. The size of the points represents the total predicted signal for a fixed laser and THz source power. We see that from the hundreds of transitions available, 4 will give us good imaging signals.}  
\label{fig:best_states}
\end{figure}

Figure~\ref{fig:best_states} shows an example of using the Monte-Carlo model to find the best THz transitions for imaging in the 0.5 - 2\,\si{\tera\hertz} frequency range. The predicted fluorescence signal is found by multiplying the FoM from the fluorescence simulation by the square of the dipole matrix element of the THz transition $\mathbf{d}_{\rm{off}\rightarrow\rm{on}}^2$. 
Although there are over 900 transitions in this frequency range, Fig.~\ref{fig:best_states} shows that most of them will produce negligible signals for imaging. 
For the transitions that will give good signal levels, we also need to consider the dipole matrix element of the 7S$_{1/2}\,\rightarrow \rm{THz}_{\rm{off}}$ transition, as this will affect the total number of atoms in the `THz off' state and hence the total number of atoms available for the THz transition. 
Assuming a constant laser power the number of atoms in this state will be proportional to the square of the dipole matrix element $\mathbf{d}_{\rm{7S}_{1/2}\rightarrow \rm{THz}_{\rm{off}}}^2$, represented by the colour of the scatter points in Fig.~\ref{fig:best_states}.
The four largest points (and hence the 4 best transitions for imaging via the 3-photon scheme in Cs) are the $n\mathrm{P}_{3/2}\rightarrow (n-1)\mathrm{D}_{5/2}$ transitions for $n=14,13,12,11$ from left to right. 
Since in general, lower powers are available from higher frequency THz sources, the higher frequency transitions will give lower overall imaging signals. However they will have advantages over imaging at lower frequencies for example gains in spatial resolution due to shorter wavelength.

Using the methodology above we can inform the choice of atomic species, atomic states and laser excitation scheme used in the experimental realisation. The next section lays out how such an experiment might be constructed and which components can be chosen to optimise performance.

\section{Experimental Setup}\label{sec:exptsetup}

\begin{figure}[hbt]
\centering
\includegraphics[width = 0.95\linewidth]{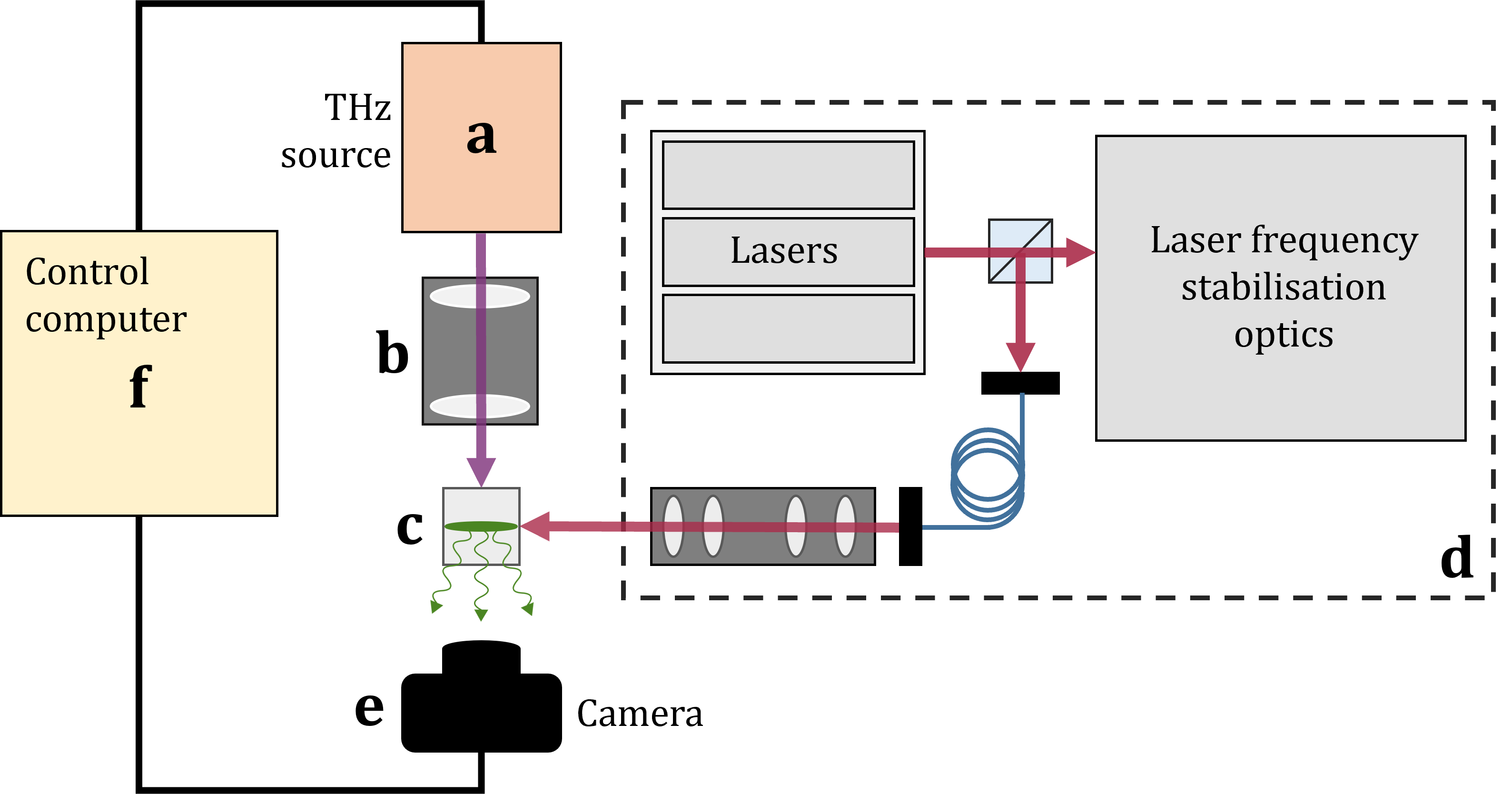}
\caption{\textbf{Basic components of an atom-based THz imaging experiment.} The key components are described in detail the the following subsections: a - THz source (see Section \ref{sec:thzsources}), b - THz optics (see Section \ref{sec:thzoptics}), c - Atomic vapour cell (see Section \ref{sec:cells}), d - Excitation lasers and optics (see Section \ref{sec:lasers}), e - Optical camera (see Section \ref{sec:cameras}), f - Experimental control (see Section \ref{sec:expt_control}).}
\label{fig:basic_layout}
\end{figure}

Here we describe the basics of a generic experimental setup to perform THz imaging in an atomic vapour cell. Figure~\ref{fig:basic_layout} shows the main components common to all vapour-based THz imaging experiments and how they are assembled to form the imaging system. Each of the key components (labelled a--f) will be described in detail in the following subsections. a - THz source (see Section \ref{sec:thzsources}), b - THz optics (see Section \ref{sec:thzoptics}), c - Atomic vapour cell (see Section \ref{sec:cells}), d - Excitation lasers and optics (see Section \ref{sec:lasers}), e - Optical camera (see Section \ref{sec:cameras}), f - Experimental control (see Section \ref{sec:expt_control}).

\subsection{Terahertz Sources}\label{sec:thzsources}

As this imaging technique involves driving resonant electric dipole transitions between Rydberg atomic states, we require a single frequency, narrowband, tunable source of coherent terahertz illumination. Although it is possible that pulsed terahertz sources could be used for atom-based imaging, most reported demonstrations have used continuous-wave (CW) terahertz sources and so we will limit the discussion to those. 
The following is not intended as a comprehensive review of terahertz sources, as that is available elsewhere e.g.~\cite{Lewis2014,Bruendermann2012}, rather a brief guide to the pros and cons of the most accessible sources from the viewpoint of researchers working on atom-based terahertz imaging.

Amongst the most convenient and accessible terahertz sources are amplifier multiplier chains (AMCs). AMCs are electronic terahertz sources based upon Schottky diodes~\cite{Maestrini2008}. They are an ideal source of terahertz illumination because they operate at room temperature, use a standard mains electrical supply and provide a fixed frequency-multiplication of a tunable seed microwave signal. Therefore the output terahertz frequency can be simply `dialed up' by changing the frequency of the microwave synthesizer used to seed the system. This can be particularly useful because the resonant frequency of the terahertz transition can often differ by 10s of MHz from the frequencies calculated by ARC due to uncertainties in the quantum defects \cite{Downes_Thesis}. The terahertz beam is typically launched into free space from a diagonal horn antenna, producing a mostly TEM$_{00}$ Gaussian mode.
Commercial products provide output power in the range of 1\,\si{\micro\watt} up to a few mW at frequencies up to around 1.1\,\si{\tera\hertz}. In general, broadly tunable models can span a few hundred GHz but produce significantly lower peak powers than narrower sources, and higher frequency sources provide significantly less output power than lower frequency models. The linewidth of the emission is typically very narrow, $<$0.1~kHz.  There are a number of commercial suppliers including Virginia Diodes~\cite{VDI} and Lytid~\cite{Lytid}.  
A broadly tunable AMC provides great versatility for terahertz imaging because it allows access to a large number of atomic transitions. 

Another convenient source of tunable terahertz radiation is the photomixer. Photomixers are based on optical heterodyning in high-bandwidth semiconductors where the output of two CW lasers is converted into terahertz radiation, exactly at the difference frequency of the lasers~\cite{Preu:2011}. The core component is a microscopic metal-semiconductor-metal structure which is irradiated with a bichromatic near-infrared field produced by the lasers. Applying a bias voltage to the metal electrodes generates a photocurrent that oscillates at the beat frequency. An antenna surrounding the photomixer emits an electromagnetic wave at the terahertz difference frequency. Commercial systems based upon compact and controllable distributed feedback (DFB) lasers are capable of producing tunable output over a very broad range of frequencies. Photomixing sources can produce up to $100\,\si{\micro\watt}$ of power at low frequencies $\sim$100~GHz but the power reduces with increasing frequency and at 500~GHz, a maximum power of around $10\,\si{\micro\watt}$ is achievable. The frequency control is achieved via temperature tuning of the lasers and the typical accuracy is around 1~GHz but can be controlled to a few MHz. The spectral linewidth is on the order of a few MHz. Commercial systems are available from Toptica Photonics~\cite{toptica_tds}.

To access frequencies above 1.5~THz, Terahertz-Quantum Cascade lasers (QCLs) are a good option. QCLs are compact semiconductor unipolar lasers where photon emission is based on electronic intersubband transitions in a cascaded quantum-well superlattice structure~\cite{Williams2007}. The characteristics of the QCL depend on the design and engineering of this semiconductor superlattice. QCLs are among the most promising terahertz sources because they can be designed to operate across a broad frequency range (from 2 to 5\,\si{\tera\hertz})~\cite{Huebers2019} with high output powers (up to 1~W)~\cite{Li2014}. However, in contrast to AMCs and photomixers, cryogenic cooling is required for operating QCLs in the terahertz range and the tuning range is much narrower, meaning that a single QCL would probably only be able to address a single terahertz imaging transition. Nowadays, different compact cooling options are commercially available including continuous Helium-flow optical cryostats \cite{Janis} and Stirling coolers \cite{Richter2010}. QCLs can be operated in CW or pulsed-mode operation, with pulses providing significantly higher peak powers. The output frequency can be tuned by changing the bath temperature in the cryostat and/or by applying a DC current to the QCL. This allows us to scan a few GHz around the engineered single-mode emission frequency~\cite{Schrottke20}.
The practical linewidth of a QCL depends on the stability of the driving current and the temperature. A QCL operated in a He-flow cryostat, free from mechanical vibrations and with a temperature stability of 2.2 mK will correspond to a linewidth of 750 kHz \cite{Wienold18}. 
If frequency stabilization is applied, the linewidth of the QCL can be reduced further to a few hundred kHz \cite{Alam19}. Several groups have demonstrated frequency locking of a THz-QCL using the center frequency of a molecular absorption line such as methanol as a frequency reference \cite{Richter2010-2,Hagelschuer:16,Ren2012}. So far most QCLs for terahertz applications have been developed in research groups (e.g. Berlin, Cambridge, Leeds). Traditionally, commercial solutions have been limited to the sub-terahertz range (wavelength below 15~\si{\micro\metre}) but recently QCLs between 2--5\,\si{\tera\hertz} have become commercially available \cite{Lytid}, with good power stability but limited frequency tunability.

Another terahertz source requiring cryogenics is the Intrinsic Josephson junction. They are based on a high T$_{\rm c}$-superconductor such as  Bi$_{2}$Sr$_{2}$CaCu$_{2}$O$_{8+\gamma}$(BSCCO) which is single crystal containing superconducting layers. Terahertz emission is based on the intrinsic Josephson effect between arrays of junctions (stacks) that naturally convert a direct-current voltage into a high-frequency electric current (1 mV corresponding to 0.4836\,\si{\tera\hertz})~\cite{Li2014}. This solid state CW THz source emits frequencies from 0.3 to 1.5\,\si{\tera\hertz} and power up 100~\si{\micro\watt}~\cite{Kleiner2019}. Above 1.5\,\si{\tera\hertz}, powers up to 10~\si{\micro\watt} have been reported~\cite{Welp2013}. Like with QCLs, phase-locking techniques can be established using the absorption spectra of water and ammonia vapour~\cite{Sun2017} for frequency stabilization.

Also worthy of mention are terahertz gas lasers which are typically excited using a CO$_2$ pump laser and have been a source of CW coherent terahertz radiation above 0.3\,\si{\tera\hertz} for many years \cite{CHANG1970423}. They are based on rotational transitions of molecules meaning that they are not continuously tunable. Gas lasers can reach up to 1~mW of THz power for 1~W of pump power. Recently it was demonstrated that the combination of a quantum cascade laser with a nitrous oxide molecular gas laser, can offer a widely tunable THz source \cite{Chevalier2019}. Finally, we mention two other terahertz sources, the first is the backward wave oscillator (BWO) which produces narrow linewidth, tunable terahertz radiation with high output power. In recent times, manufacture of these devices has all but ceased as AMC sources have become higher power and provide a more compact and versatile solution. Finally we mention the free-electron laser (FEL), which is a facility-scale radiation source. FELs can produce very high (several watts) output power in the terahertz range. A summary of the relevant properties of various terahertz sources is shown in Table~\ref{tab:thz-sources}. 

Once the terahertz source is installed within the set up, we require a system of lenses and mirrors to direct the terahertz beam to the target and then image the target onto the atomic vapour. The next section discusses the terahertz optics and materials required to achieve this.



\begin{table}
\begin{center}
\resizebox{\textwidth}{!}{%
\begin{tabular}{ |c |c |c |c |c|}
\hline
Source & Frequency [THz] &  Tunability [GHz] & Power [mW]  & Linewidth [kHz] \\  [0.5ex] 
\hline
Amplifier Multiplier Chain & 0.1 -- 1.1 & $\approx$ 250 & 0.01 & $<$~0.1  \\  
THz-QCL & 2.5 -- 5.6  \cite{Huebers2019} & 3 -- 10  \cite{Lytid,Bruendermann2012}& 10 -- 1000 \cite{Bruendermann2012,Li2014}& 750~\cite{Wienold18}\\
Josephson emitter & 0.4 -- 2.7 \cite{Welp2013} & 50 -- 70 \cite{Kashiwagi2015} & $<$ 0.1 \cite{Welp2013,Kleiner2019} & 700 -- 1000 \cite{Sun2017,Kleiner2019}\\ 
Photomixer & 0.1 -- 1.5 & $>$~1000 & 0.01 @ 500~GHz & $\approx$ 2000 \\ 
THz Gas laser (data from \cite{Bruendermann2012}) & 0.5 -- 5.0 & - &1 & 1000  \\ 
BWO (data from \cite{Bruendermann2012})& $<$~1 & 10 -- 140 & 1 -- 10 & 10000\\ 
Free electron laser (data from \cite{Bruendermann2012}) & 1 -- 10 & - & 10000 & -\\ \hline
\end{tabular}}
\caption{Frequency range, tunability, power and linewidth of different terahertz sources. }
\label{tab:thz-sources}
\end{center}
\end{table}

\subsection{Terahertz Materials and Optics}\label{sec:thzoptics}

\begin{figure}[t]
    \centering
    \includegraphics[width = 0.98\linewidth]{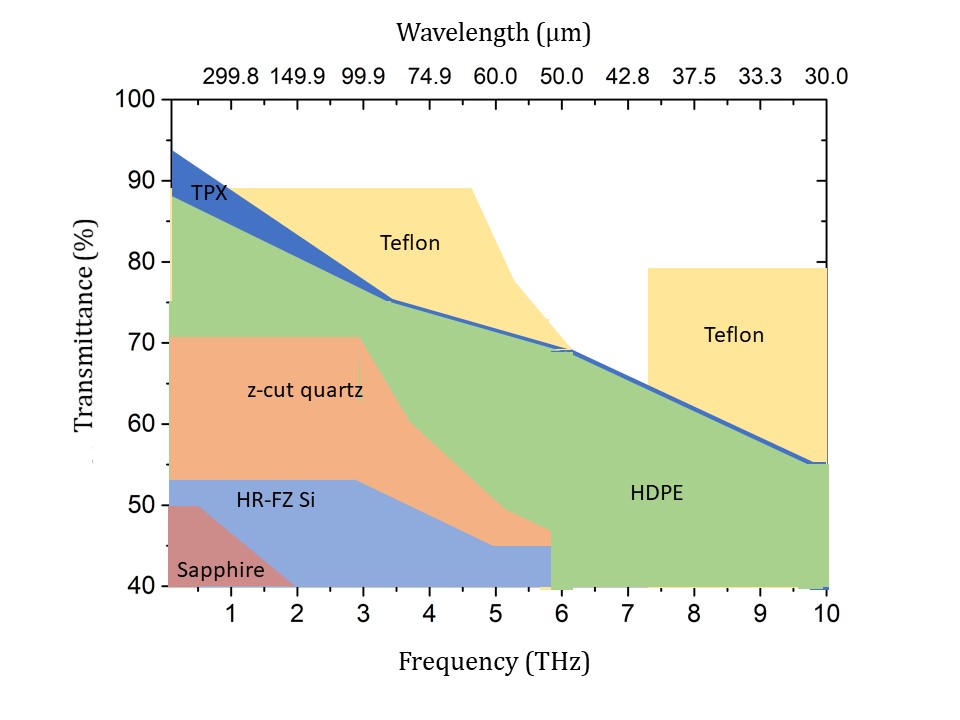}
    \caption{\textbf{Transmittance of common terahertz materials.} The data is shown for substrates with a thickness of 1-2 mm for $z$-cut quartz, Sapphire, TPX while HDPE, HR-FZ Silicon and Teflon have thickness 3~mm, 5~mm and 0.1~mm respectively (data from \cite{Tydex}).}
    \label{fig:transmission}
\end{figure}

As with any optical system, terahertz systems require elements such as mirrors, lenses, beamsplitters and polarisers. Most of the ordinary glasses commonly used in the optical region are useless for terahertz applications as they are opaque at these wavelengths and so, as terahertz technology has evolved, the development of transmissive materials suitable for lenses and windows \cite{Islam2020,Rogalin2018,Tydex} has become imperative. 
When terahertz radiation passes through a material, it can be partly reflected at the front surface with coefficient $R$, and absorbed by the material with coefficient $A$, causing radiation losses. The reflectivity from a surface depends upon the angle of incidence and upon the plane of polarization of the light. The general expression for reflectivity is derivable from Fresnel's equations and the reflection loss $R$ per surface at normal incidence is dependent upon the indices of refraction of the two media as

\begin{equation}
\displaystyle R=\left|{\frac {n_{1}-n_{2}}{n_{1}+n_{2}}}\right|^{2}\, ,
\end{equation}
where the refractive index of air $n_1 \approx 1$ and $n_2$ is the index of refraction of the material used.
The reflection coefficients can be used to calculate the amount of light transmitted ($T=1-R$) through a medium if the only losses are reflection losses, i.e. $A=0$. However, in general, losses through an optical component will be caused by reflection due to refractive index mismatch but also absorption though the material. Often the `transmittance' of a component will be quoted, which depends on the substrate thickness as it is the sum of transmission and reflection losses. Organic materials such as polymers can be excellent transmitters in the terahertz range with transmission values between 70\%--98\% quoted for wavelengths between 200~\si{\micro\metre} and 3000~\si{\micro\metre} and material thicknesses of 3--0.1~mm. Some recently-developed polymer materials include poly-4-methypentene-1 (TPX), high-density polyethylene (HDPE), Zeonex and polytetrafluorenthylene (teflon). For a long time, the most commonly used material in terahertz optics has been high-resistivity silicon grown by the floating-zone technique (HR-FZ Si) with a transmittance of 50-54\% for a 5~mm thickness substrate starting from minimum wavelengths around 50~\si{\micro\metre}. Some crystalline materials have relatively high transmission in the terahertz range. For example a 2~mm thick sample of $z$-cut quartz ($\alpha$) has $\>$65\% transmission at wavelengths above 50~\si{\micro\metre} \cite{Naftaly2021} and Sapphire has around 45\% transmission for a 2~mm sample at wavelengths above 2000~\si{\micro\metre}. Furthermore, $z$-cut quartz and TPX are often used in terahertz systems because they are relatively transparent across both the terahertz and visible range, making them very practical for alignment purposes. Table~\ref{tab:materials} summarises common terahertz materials, their refractive indices and reflection losses $R$ per surface and Figure~\ref{fig:transmission} summarises the transmittance of a number of terahertz materials at a given thickness which can be used as a quick reference.

\begin{table}[hbt]
\begin{center}

\begin{tabular}{ |c | c| c|}
\hline
Material& $n$ & R (\%) \\ \hline
Zeonex & 1.529 &  1.31\\ 
TPX & 1.455 &  3.43 \\ 
HDPE & 1.535 & 4.45  \\ 
PTFE, Teflon & 1.466 & 3.57 \\ 
Picarin & 1.56 & 4.78 \\ 
PET, Mylar & 1.72 & 7.00  \\ 
Fused Silica/quartz & 1.943 & 10.27 \\ 
BK7 & 2.46 & 17.805\\ 
Duran/borosilicate & 2.06 & 12.0  \\ 
$z$-cut quartz crystal & 2.1 ($n_e$) 2.6 ($n_o$) & 12.59 ($n_e$) 19.75 ($n_o$)  \\ 
HRFZ Silicon & 3.416 & 29.93 \\ 
Sapphire & 3.6 ($n_e$) 3.1 ($n_o$) & 31.94 ($n_e$) 26.23 ($n_o$)  \\ [1ex]  \hline
\end{tabular}
\caption{Refractive index ($n$) and reflection coefficients $R$ for common materials. Data from \cite{Rogalin2018, Islam2020, Bruendermann2012}. Quartz and sapphire crystals are birefringent and so the ordinary ($n_o$) and extraordinary ($n_e$) refractive indices are both presented.}
\label{tab:materials}
\end{center}
\end{table}

The first optic in most terahertz systems will typically be a collimating optic as the output of most terahertz sources is highly divergent. There are two options to achieve this; either with plano-convex lenses made of teflon, HRFZ Si or TPX that are commercially available from \cite{batop,Swiss,menlo,terasense,Thorlabslens,Tydex,Lytid} or by using an off-axis gold coated parabolic mirror which comes in several focal lengths and large diameters \cite{Lytid,thorlabs}. It should be mentioned that if one needs to use a short-focal length lens or parabolic mirror for collimation, it is better to use a gold-coated parabolic mirror over a transmissive lens. Even if the lenses are made of a material with a high terahertz transmittance, the lens will need to be thick and power losses could become significant. As a general rule, the number of transmissive components should be minimised to prevent attenuation of the terahertz signal through the imaging system and reflective optics using polished metal surfaces (typically reflectivities $>$ 98\%) should be used instead. However, some transmissive optics are essential, such as the atomic vapour cell. Also, as discussed above, some terahertz sources need to be cooled in cryostats and windows are required to hold the vacuum while transmitting the terahertz radiation with as little power loss as possible. Several companies provide suitable cryostat windows for terahertz radiation \cite{Lytid,Tydex,QMC}.
Due to the absorption of terahertz radiation by atmospheric water vapour, the optical path often needs to be designed as short as possible (ideally less than 0.5~m) or alternatively it needs to be evacuated by enclosing the system or kept in positive pressure using dry nitrogen to reduce humidity. 

Other required components may include 50:50 beam splitters, where a simple solution is to use a 5~\si{\micro\metre} thick film of polyethylene terephthalate (Mylar). Usually in order to cover the whole terahertz range, several Mylar beam splitters with different thicknesses need to be used. An alternative is to use a silicon beam splitter, several millimeters thick, where the entire terahertz range can be covered with a single optic~\cite{Homes:07}. Commercial silicon beam splitters are also available \cite{Tydex,Lytid} that have been shown to be identical to Mylar at 3.4\,\si{\tera\hertz} \cite{Wieland2022}. Also, dichroic beamsplitters may be required to combine terahertz and visible or NIR wavelengths and these can be also found commercially~\cite{QMC,newport,Lytid}. Components such as polarizers, filters and attenuators are nowadays as well developed in the terahertz region as they are for optical frequencies. In order to control the polarization of the terahertz field, homemade solutions such as sheets of paper have been demonstrated~\cite{Scherger:11} alongside commercial options for $\lambda/4$ and $\lambda/2$ waveplates available from numerous suppliers~\cite{purewave,hammatsus,Tydex,anteral,Lytid}. 
Wire grid polarisers suitable for frequencies below 3~THz are also available from QMC~\cite{QMC}. Terahertz filters can be used to remove all radiation up to a predetermined wavelength and then transmit at longer wavelengths, or in a delimited band. These are available from several commercial sources~\cite{Tydex,VDI, QMC,Lytid}. Power attenuators can also be found at several companies \cite{Lytid,Tydex}. Finally, terahertz radiation does not penetrate into metals and a good absorber is the CR-series of Eccosorb for use as a terahertz blocker~\cite{ Bruendermann2012}.

Further information covering terahertz materials, optics and technology in general is presented in this book \cite{Bruendermann2012}. 

\subsection{Atomic Vapour Cells}\label{sec:cells}

The atomic vapour cell is an essential component of the experimental setup. 
Cells require good optical access for the excitation lasers and visible fluorescence, but also need to allow the terahertz field to pass through without significant loss or distortion. Reflection and absorption at the interface can be made to be small by choosing suitable materials. Depending on the frequency range of the terahertz field, different materials should be considered for cell fabrication. 

For the low-terahertz range, below $\sim$1.1\,\si{\tera\hertz}, standard glass cells are safe to use as power transmittance is between 60-80\% for a window of 1-2 mm thickness \cite{Tydex} and a good choice is UV-fused silica (quartz glass) over Duran (Schott borosilicate) glass.
For the high-terahertz range, above $\sim$1.1\,\si{\tera\hertz}, standard glass vapour cells cannot be used due to the large reflection and absorption losses. Multiple internal reflections can occur within the vapour cell as the reflective parallel windows behave as a Fabry-Perot (FP) cavity causing internal standing waves which can significantly reduce the sensitivity and distort the imaging. 
Alternative materials which are both transparent for visible and near-IR radiation and that satisfy the condition of low absorption for the terahertz range are required. It should also be mentioned that the effects of the vapour cell geometry on the measured terahertz field, such as the etalon effect mentioned above can be reduced with the help of using tilted windows or by making the cell size $D$ small in comparison to the wavelength $\lambda_{\rm THz}$ of the terahertz radiation, i.e  $\frac{D}{\lambda_{\rm THz}} < 0.1 $ \cite{Fan2015}. For terahertz waves, this will mean a cell size of 10--50~\si{\micro\metre}. Furthermore, in this regime, wall effects will play a bigger role and should be considered. Some alternative solutions to allow a small interaction region whilst avoiding surface interactions were suggested by {\v{S}}ibali{\'c} \textit{et al.} in \cite{sibalic2016}. Due to difficulties in cell manufacturing and optical access for micrometer scale cells, the cells used in reported experiments are on the millimeter~\cite{Wade2017} and centimeter~\cite{Downes2020} scale.

For caesium or rubidium vapour cells, we require a high number of atoms held under vacuum in order to avoid chemical reactions with the oxygen or water in the atmosphere. 
There are several options for cell bonding including the use of low-outgassing epoxy glues (Epo-Tek 377, TorrSeal) and different glass bonding techniques such as thermal fusing, glass-to-glass bonding and anodic bonding~\cite{Daschner14}. Using $z$-cut quartz windows and a borosilicate/quartz cell body proved to be unsuccessful due to the fact that the thermal expansion of $z$-cut quartz is highly anisotropic and its thermal expansion coefficients~\cite{uqgoptics} differ significantly from other glasses. This mismatch causes large internal stress at high temperatures leading to breakage during thermal bonding techniques. The option of gluing the window to the cell body was not optimum due to the deterioration of the vacuum caused by chemical reactions between rubidium and the epoxy, as reported by Daschner~\cite{Daschner2015}.
The most suitable solution was found to be using HR-FZ-silicon windows~\cite{korth} or HR-FZ-silicon wafers~\cite{topsil} for the terahertz radiation and a borosilicate window for the lasers which excite the atoms into Rydberg states~\cite{Torralbo2022}.  
As was discussed in Section~\ref{sec:thzoptics}, the terahertz transmittance of silicon is not high due to high reflection losses. However, reflection losses can be reduced by the use of a terahertz antireflection (AR) coating.
Parylene coating technology for plane surfaces is well known in microelectronics and it has has been shown that a double-side AR thin film coating with Parylene on terahertz windows and lenses can increase the transmittance up to 90$\%$ \cite{Tydex,HUBERS200141}. 
Silicon-glass substrates are excellent materials for the anodic bonding sealing technique.
It should be mentioned that anodic bonding requires very flat and highly polished surfaces ($\lambda$/10 at 633 nm) and substrates may need to be treated ahead of the fabrication process \cite{korth}.  
Once the empty cell is fabricated, it is connected to a vacuum manifold and filled with atomic vapour by breaking and then heating an ampule containing the chosen alkali-metal. After filling the reservoir with alkali metal, the cell is fully sealed using a gas flame. Once manufactured and sealed, vapour cells will last for many years and do not need replacing. 

In the imaging experiments the atomic vapour cell is mounted in an oven which maintains the optical access. For the case of caesium, we heat moderately to around $40\,\si{\celsius}$ and for rubidium we heat up to $60\,\si{\celsius}$. To prevent atomic vapour condensing on the inner walls of the cell and restricting optical access the reservoir (usually at the end of a glass stem) should be kept as the coolest part of the cell. The vapour cell can be initially characterised by obtaining the absorption spectrum using saturation absorption spectroscopy and EIT spectroscopy. The temperature of the vapour can be found by fitting the measured weak-probe absorption spectrum to theory using the ElecSus program \cite{zentile2015}.
 
\subsection{Excitation Lasers}\label{sec:lasers}

To perform the resonant excitation of atoms to high-lying states, we require tunable, single-frequency lasers with linewidths comparable to the linewidths of the atomic transitions. 
As shown previously in Figure~\ref{fig:schemes}, there are many possible ways to excite atoms to Rydberg states, but we will restrict our consideration to schemes where the laser wavelengths are easily available. This means we will consider 2- and 3-step excitation schemes. Some common pathways and the relevant laser technologies are listed in Table~\ref{table:lasers}. 

\begin{table}[hbt]
\begin{center}
\resizebox{\textwidth}{!}{%
\begin{tabular}{ |c |c |c |c |c| c|}
\hline
Scheme & Atom &  Transition  & Wavelength & Laser & Typ. Power\\ \hline
2-photon & Cs & $6\mathrm{S}_{1/2}\rightarrow 6\mathrm{P}_{3/2}$ & 852~nm & diode, DFB, Ti:Sapph & $\sim$ 5~mW\\  
&  & $6\mathrm{P}_{3/2}\rightarrow n\mathrm{S}$ or n$\mathrm{D}$ & 512 - 604~nm & diode, SHG & $> 100$~mW\\  \hline
& Rb & $5\mathrm{S}_{1/2}\rightarrow 5\mathrm{P}_{3/2}$ & 780~nm & diode, DFB, Ti:Sapph & $\sim$~5~mW\\
&  & $5\mathrm{P}_{3/2}\rightarrow$ nS or nD & 480 - 540~nm & diode, DFB, SHG & $> 100$~mW\\ \hline
3-photon & Cs & $6\mathrm{S}_{1/2}\rightarrow 6\mathrm{P}_{3/2}$ & 852~nm & diode, DFB, Ti:Sapph & $\sim 5\,\si{\milli\watt}$\\  
&  & $6\mathrm{P}_{3/2}\rightarrow 7\mathrm{S}_{1/2}$ & 1470 nm & diode, DFB  & $\sim 20\,\si{\milli\watt}$\\ 
&  & $7\mathrm{S}_{1/2}\rightarrow n\mathrm{P}$ & 790 - 980 nm & diode, DFB, Ti:Sapph & $\sim 100\,\si{\milli\watt}$\\ \hline
& Rb & $5\mathrm{S}_{1/2}\rightarrow 5\mathrm{P}_{3/2}$ & 780 nm & diode, DFB, Ti:Sapph & $\sim$ 5~mW\\  
& & $5\mathrm{P}_{3/2}\rightarrow 6\mathrm{S}_{1/2}$ & 1367 nm & diode, DFB & $\sim$20~mW\\ 
&  & $6\mathrm{S}_{1/2}\rightarrow n\mathrm{P}$ & 750 - 870 nm & diode, DFB, Ti:Sapph & $\sim 100\,\si{\milli\watt}$\\ \hline
\end{tabular}}
\caption{Examples of common Rydberg excitation schemes in Cs and Rb and the corresponding laser technology available.}
\label{table:lasers}
\end{center}
\end{table}

\subsubsection{Frequency Stabilization: }

Since the atomic transitions are narrowband, the excitation lasers need to be frequency stabilised to within a few MHz of the relevant transition. Frequency stabilisation is a common experimental technique in atomic physics experiments and so many schemes exist to enable the lasers' frequency to be `locked' to the right value. For transitions out of the ground state we can stabilise the frequency to a sub-Doppler feature in an atomic spectrum using techniques such as saturated absorption spectroscopy, polarisation spectroscopy \cite{Pearman:2002}, modulation transfer spectroscopy~\cite{McCarron_2008} etc.
In the 3-photon schemes the intermediate step laser can also be locked to an atomic reference by using excited-state polarisation spectroscopy \cite{Carr:2012, Almuhawish:2021} or double-resonance optical pumping (DROP)~\cite{Moon:07}. 
For the final laser in the excitation pathway it is still possible to use an absolute atomic frequency reference for example by using EIT locking \cite{abel:2009,Carr:12a}, but as more power is required to generate an atomic signal, this requires recycling of the light or else the power available for the main imaging experiment is reduced. However, other techniques enable the final laser to be locked to an arbitrary frequency without needing to address the Rydberg transition. 
An external frequency reference such as an ultra-low expansion (ULE) cavity provides a very stable frequency reference to which lasers can be locked using Pound-Drever-Hall (PDH) technique \cite{Fox:2003}, however they are expensive and require being held under vacuum making the experimental requirements complex. Alternatives involve a transfer cavity scheme \cite{burke:2005,Subhankar:2019} whereby the length of a tunable cavity is stabilised to a reference laser, which is in turn stabilised to an atomic reference. A second laser can then be locked to the stabilised cavity at any arbitrary frequency. Another option is to shift a ground-state atomic resonance using a large magnetic field such that it is at the correct frequency for an excited state transition \cite{Reed:2018}. 

\subsubsection{Beam Shaping: }

Once the laser frequencies are fixed at the correct value, the beams need to be shaped and directed into the atomic vapour cell. In order to create a sheet of Rydberg atoms the laser beams need to address the same region within the vapour cell, so all beams need to be overlapped. This can be achieved by combining beams using dichroic mirrors or by having beams counterpropagating through the vapour cell. To create a thin sheet of Rydberg atoms the excitation beams need to be shaped such that they form a light sheet at the position of the vapour cell. Again there are multiple different ways to accomplish this including using combinations of spherical and cylindrical lenses in telescopes to change the beam diameter in each axis or by using a commercially available fibre-coupled laser line generator such as those manufactured by Sch\"{a}fter \& Kirchhoff. 
Ideally the light sheet should be uniform as it propagates through the vapour cell, thereby reducing variations in fluorescence intensity in the optical images. 

Another key parameter is the laser intensities used to excite the atoms to Rydberg levels. Too little power and not many atoms will be promoted to a Rydberg state so the eventual fluorescence signal will be very weak, too much laser power and the atomic states can start to exhibit Autler-Townes splitting \cite{autlertownes1955,anisimov2011} which reduces the fluorescence signal. The optimum values depend on many different parameters for example excitation scheme, cell temperature/atom number density and beam diameter, so we will not quote any specific values here. 

\subsection{Camera and Data Readout}\label{sec:cameras}

The final element required in this terahertz imaging system is a method for recording the fluorescence from the atomic vapour, namely an optical camera. The requirements on the performance of the optical camera are very much dependent on the application for which the images are being captured. For instance if high-resolution low-noise images are required then a scientific EMCCD or sCMOS camera can be used. If the application requires high-speed video then a high-speed camera such as a Photron FASTCAM or Phantom TMX 7510 would be more suitable.

Images captured by the camera generally require little or no processing. The user may wish to subtract a background image (an image of the fluorescence from the vapour in the absence of the terahertz field) to get a better idea of the true effect of the terahertz field on the vapour and reduce background noise. Since the resolution of optical cameras is often greater than the resolution of the vapour, optical camera pixels can be binned to increase the signal to noise ratio without any loss of spatial information.

\begin{figure}[t]
\centering
\includegraphics[width=0.9\textwidth]{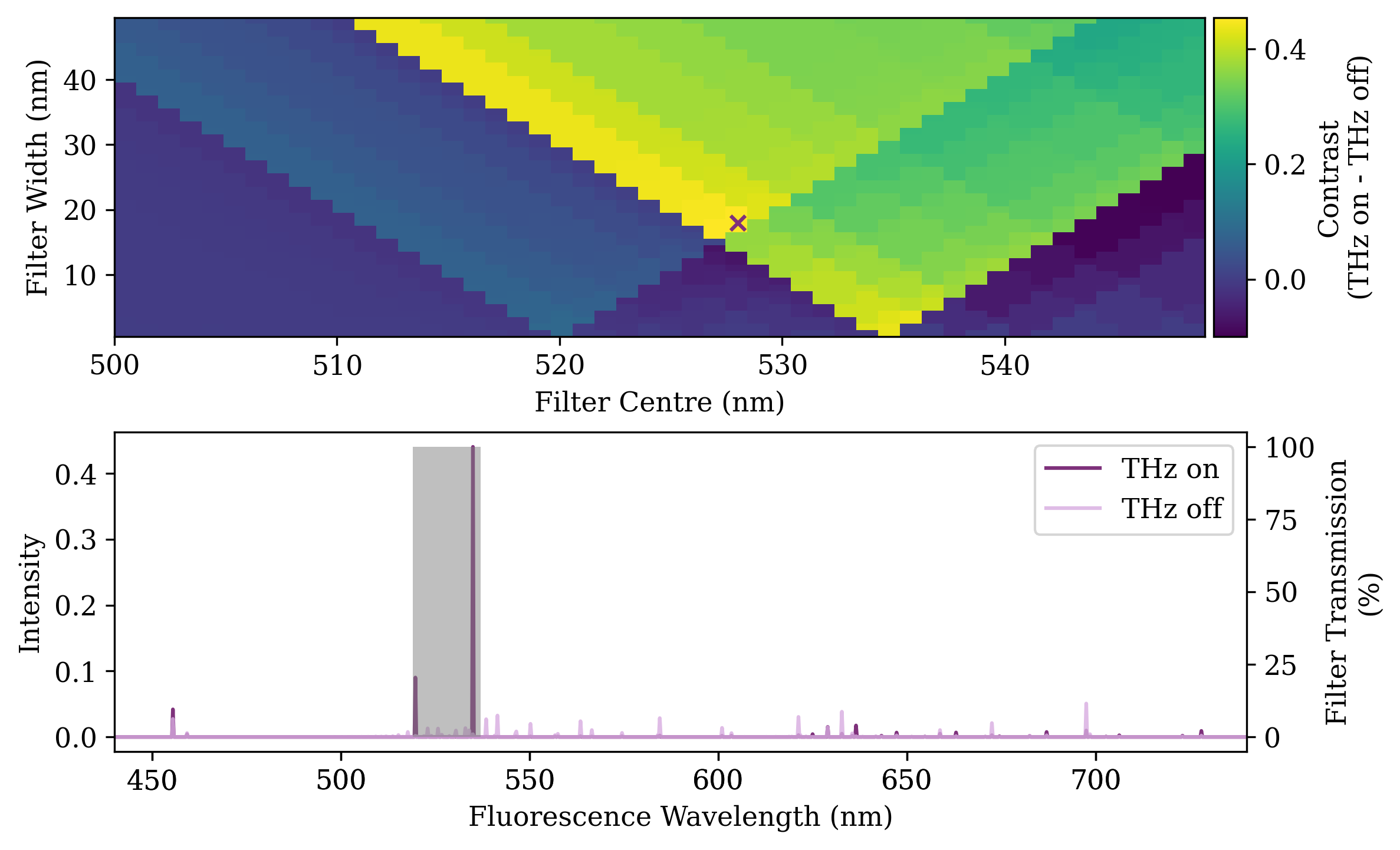}
\caption{\textbf{Selecting an optical filter.} Filter selection for terahertz imaging using the $14\mathrm{P}_{3/2}\rightarrow13\mathrm{D}_{5/2}$ transition in Cs. Top: Evaluating the contrast figure of merit ($\mathrm{THz}_{\mathrm{on}} - \mathrm{THz}_{\mathrm{off}}$) for a range of filter widths and centre wavelengths. The cross shows the location of a filter leading to maximised contrast (17\,\si{\nano\metre} width, 528\,\si{\nano\metre} centre). Bottom: An example of the fluorescence spectra in the terahertz off and terahertz on cases (light and dark purple lines respectively) with the optimum filter transmission window shown by the grey shaded region.}
\label{fig:filter}
\end{figure}

If a 3-step excitation is used, the fluorescence signal will be visible to the naked eye as the IR excitation lasers are beyond the normal optical response. 
However optical cameras are often sensitive in the near infrared, so in this case a low-pass filter needs to be applied to block out scattered light from the excitation lasers. If true-colour images of the vapour are required this could simply be a reasonably cheap low-pass filter with a cut-off wavelength in the NIR range, below the wavelength of the final step laser (for example a Thorlabs FES0750 or Newport 10SWF-750-B). If spectral information about the vapour is not required then a narrower band-pass filter can be chosen that maximises the contrast between the the terahertz on and off cases. We can once again use the Monte Carlo model of emitted fluorescence to choose the optimum filter width and centre wavelength to give maximum signal. This gives a range of parameters to search for a suitable product, for example from Semrock. Figure~\ref{fig:filter} shows an example of using the Monte Carlo model to evaluate the optimum filter parameters for imaging using the $14\mathrm{P}_{3/2}\rightarrow13\mathrm{D}_{5/2}$ transition in Cs. From this we can see that the best choice of filter will be one that allows fluorescence emitted between 520\,\si{\nano\metre} and 535\,\si{\nano\metre} to be transmitted, for example a $17\,\si{\nano\metre}$ wide filter centred on $528\,\si{\nano\metre}$.
In the case of using a 2-step excitation, a filter will need to be chosen that isolates the visible fluorescence from the scatter due to the visible excitation lasers. 

\subsection{Experiment Control}\label{sec:expt_control}

Since the imaging scheme described here is continuous wave, there are no complex timing and control sequences that need to be implemented in order to capture images. As long as the excitation lasers remain locked at the correct frequency, a terahertz image can be observed in the vapour cell and captured on a camera. In cases where an image in the absence of the terahertz field is required for background subtraction, the terahertz field and camera can be easily switched using a computer and experimental control software such as LabView or Python. Since the vapour response is on the order of microseconds there is no requirement to wait for the system to `recover' between shots. Using a software package that enables a live view of the feed from the camera (such as MicroManager) means that the image from the vapour can be used to aid alignment. 

\section{Results and Characterisation of Performance}\label{sec:results}

In order to explore the performance of an atom-based terahertz imaging system we use as an example an experiment based on Cs vapour, as discussed in \cite{Downes2020, Downes_Thesis}. In this particular experiment a 3-photon excitation is used to reach an initial Rydberg state ($14\mathrm{P}_{3/2}$), from which a resonant 0.5~THz field can transfer population to a nearby state ($13\mathrm{D}_{5/2}$). 
While the exact numbers will vary depending on the atomic species and states used, the principles of the system's characterisation will remain the same between experiments.

\subsection{Spatial Resolution}\label{sec:resolution}

When considering the spatial resolution of this imaging technique there are two main aspects to consider; namely the resolution of the vapour and the resolution of the entire terahertz `imaging system' comprising vapour detector and terahertz imaging optics. 
We assume that the ultimate limit on the resolution of the vapour will be set by the motional blurring effect arising from the fact that the atoms will move a finite amount between the time in which they interact with the terahertz field and the time at which they decay and emit a signal photon. 
Assuming the speed of the atoms is well represented by the Maxwell-Boltzmann distribution of speeds, the most probable speed of an atom in a vapour at 50\si{\celsius} is approximately $200\,\si{\metre\second}^{-1}$. The low-lying Rydberg states used here have lifetimes on the order of $800\,\si{\nano\second}$ \cite{Sibalic2017} meaning that on average an atom travels $160\,\si{\micro\metre}$ between absorbing a terahertz photon and emitting a green photon, placing a lower bound on the resolution of the vapour itself. This limit is around 4 times smaller than the wavelength of the terahertz radiation meaning that the vapour can be used to image sub-wavelength structures in the near-field. 
\begin{figure}[t]
\centering
\includegraphics[width=0.9\textwidth]{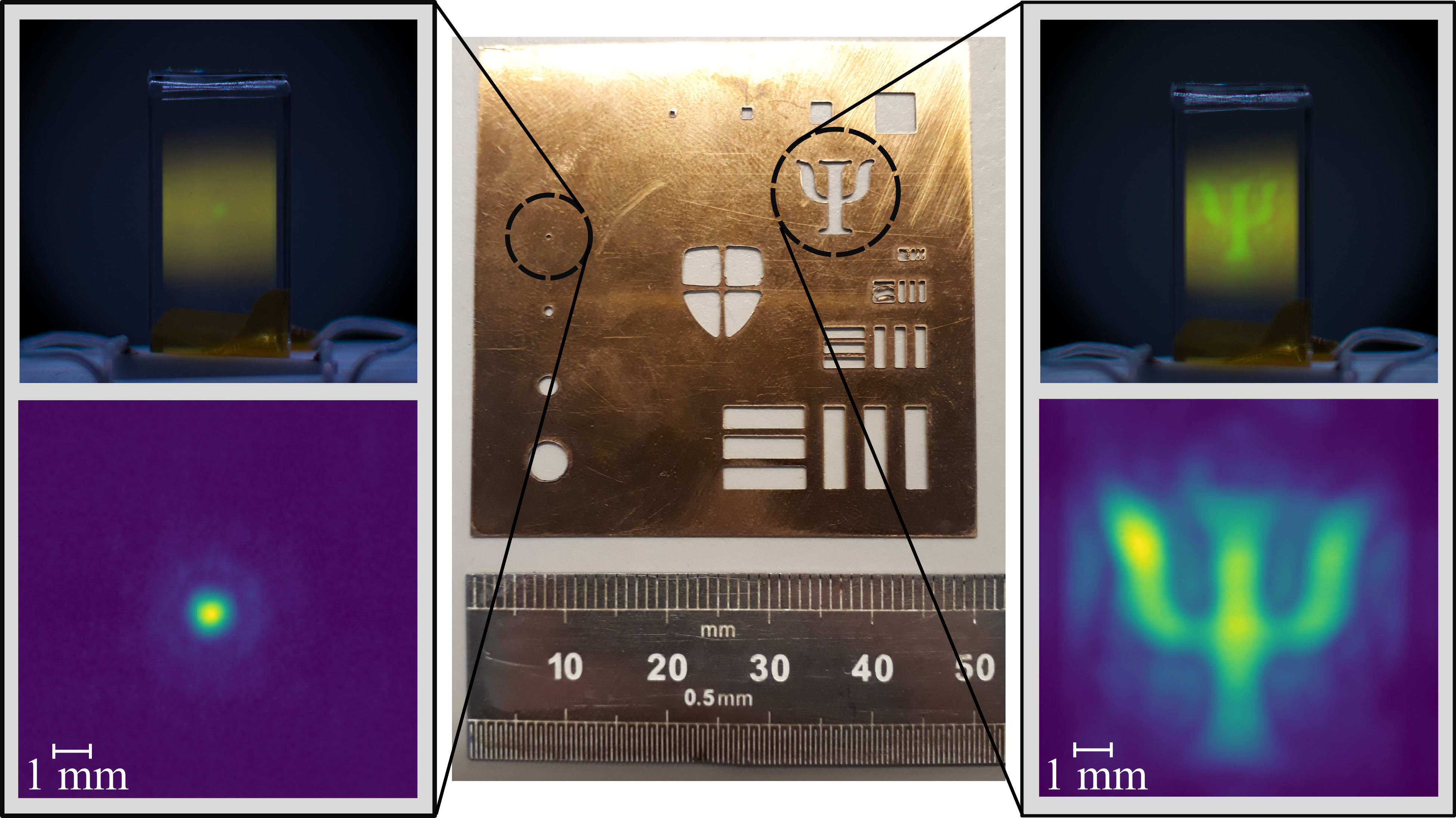}
\caption[Demonstration of spatial resolution]{\textbf{Demonstration of spatial resolution.} A metal mask (centre) is placed in the object plane of the system. To the left and right are true colour images taken with a DSLR camera (above) and false colour images taken with an Andor iXon (below) for a 0.50\,\si{\milli\metre} diameter pinhole (left) and a `$\Psi$'-shaped aperture (right). Figure taken from Downes \textit{et al.} Phys. Rev. X. \textbf{10}, 011027  (2020) \cite{Downes2020}.}
\label{fig:spatial}
\end{figure}

\begin{figure}[t]
\centering
\includegraphics[width=0.95\textwidth]{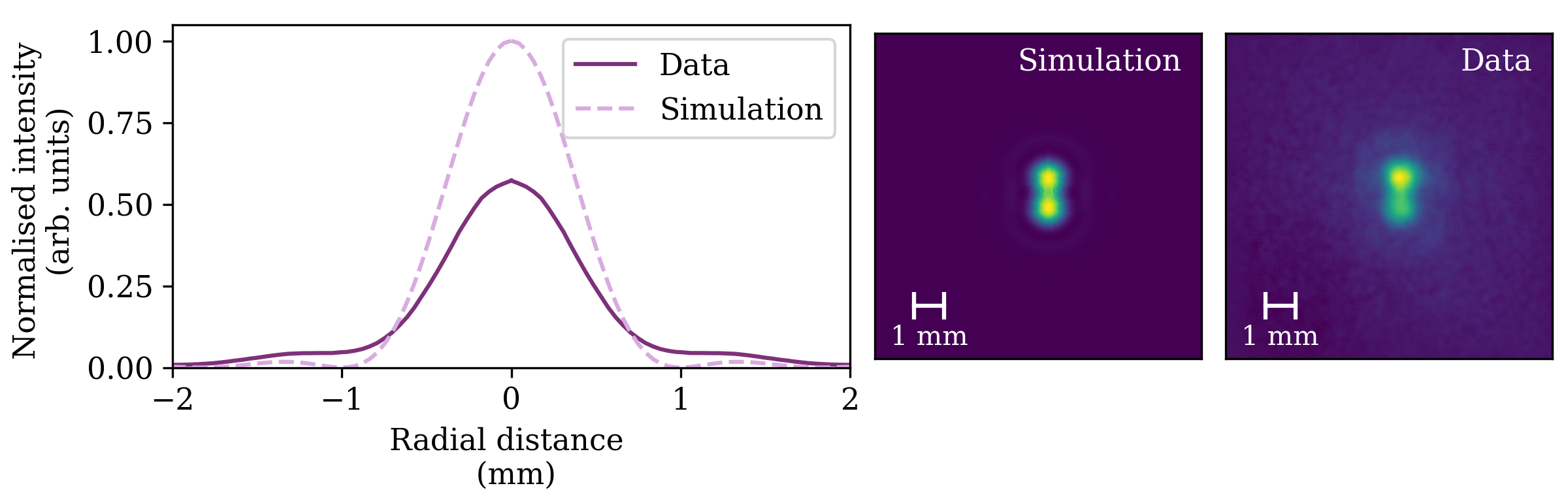}
\caption{\textbf{Measuring spatial resolution via the Rayleigh criterion.} Radial average of a single pinhole image (dark purple solid line) is compared to the diffraction limited case (light purple dashed line). The data and simulation are normalised to have the same integrated power. This simulated Airy pattern is then used to predict the results of imaging two sub-wavelength pinholes separated by 1\,mm (centre). The pinholes are resolved in the image (right).}
\label{fig:Airy}
\end{figure}

A different measure of the resolution of the system is to consider the system's ability to image structures in the far field. 
In order to do this it is helpful to consider the Rayleigh criterion, that two Airy patterns of equal intensity are `just resolved' when the maximum of one lies over the first minimum of the other \cite{F2F}. 
An image of a point source by a perfect imaging system will be given by an Airy pattern, in which the intensity $I$ at a radial distance $r$ is described by
\begin{equation}
    I(a) = I_0\left(\frac{2J_1(a)}{a}\right)^2,
    \label{eqn:Airy}
\end{equation}
where $J_1(x)$ is the Bessel function of the first kind and
\begin{equation}
    a = \frac{\pi r}{\lambda N}.
\end{equation}
Here $\lambda$ is the wavelength of the imaging light, and $N$ is the f--number of the imaging system.
The factor $I_0$ defines the maximum intensity at the centre of the image. 
For the imaging system demonstrated here the f--number is estimated to be 1.5, and the imaging wavelength is 0.55\,\si{\milli\metre}. This gives a theoretical resolution of $1.0\,\si{\milli\metre}$, meaning the system should be able to resolve two point sources separated by $1.0\,\si{\milli\metre}$. This is demonstrated in Fig.~\ref{fig:Airy}, in which we show an image of two sub-wavelength diameter pinholes separated by $1.0\,\si{\milli\metre}$. 


\begin{figure}[t]
    \centering
    \includegraphics[width = 0.9\linewidth]{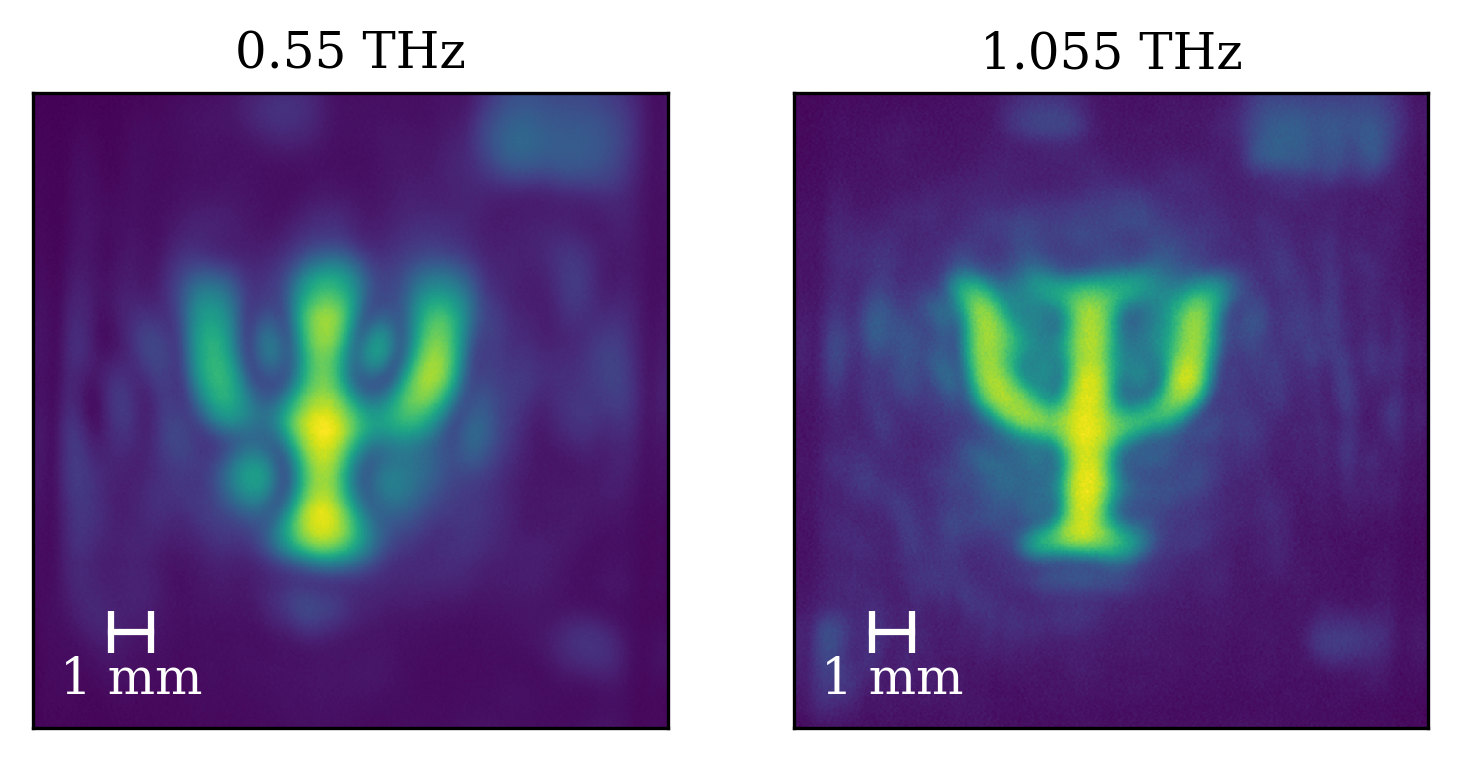}
    \caption{\textbf{Spatial resolution improvement with reduced wavelength.} The same `$\Psi$'-shaped aperture as shown in Fig.~\ref{fig:spatial} is imaged using the same imaging system but at two different imaging frequencies/wavelengths. The image taken at $1.055\,\si{\tera\hertz}$ (right) shows an improvement in resolution when compared to the image taken at $0.55\,\si{\tera\hertz}$ (left).}
    \label{fig:resolution}
\end{figure}

From the expression defining $a$ we can see that the value of $r$ at which the first minimum of the Bessel function occurs is inversely proportional to the wavelength of the imaging light, meaning for instance that a factor of 2 reduction in wavelength will lead to a factor of 2 improvement in resolution, assuming the same lens system is used. This resolution improvement is illustrated in Fig.~\ref{fig:resolution}. An image of the `$\Psi$'-shaped aperture from the mask in Fig.~\ref{fig:spatial} is imaged at two different imaging wavelengths using the same lens system. It can be seen that the reduction in wavelength when moving from $0.55\,\si{\tera\hertz}$ ($\lambda = 550\,\si{\micro\metre})$ to $1.055\,\si{\tera\hertz}$ ($\lambda = 280\,\si{\micro\metre}$) results in a sharpening of the image despite the fact that no other parameters have changed. 

\subsection{Temporal Resolution}\label{sec:temporal}
\begin{figure}[ht]
    \centering
    \includegraphics[width = 0.95\linewidth]{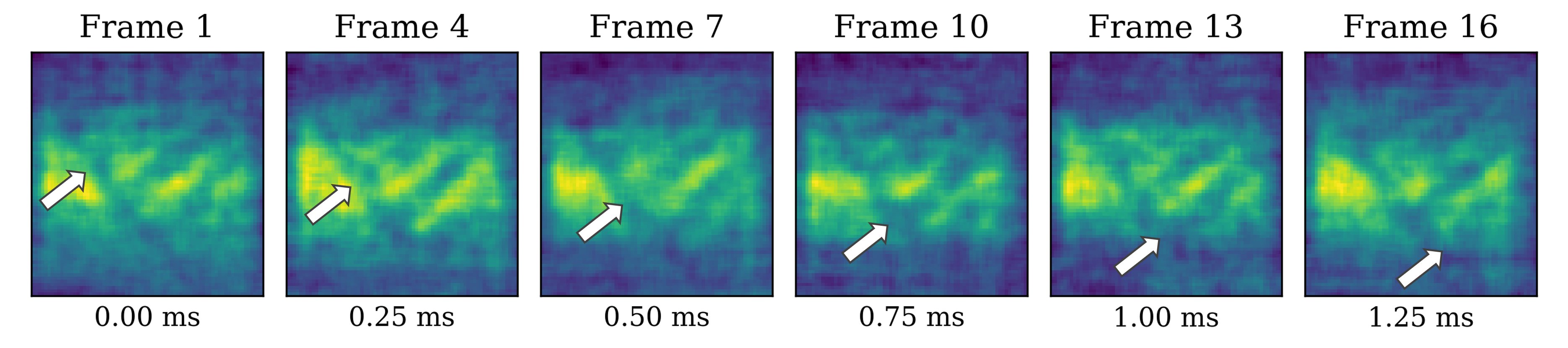}
    \caption{\textbf{Frames from a 12k fps terahertz video.} Static frames from a terahertz video of a rotating optical chopper wheel taken at 12,000 frames per second. Only every third frame is shown here and the arrows have been added to highlight the movement of one spoke of the wheel between frames. The images were taken using a Phantom TMX 7510 camera and have undergone 4x4 binning and smoothing to improve clarity.}
    \label{fig:chopper}
\end{figure}

Since this atom-based imaging scheme relies on the time taken for excited atoms to decay and emit a visible photon it can capture images incredibly fast. For low-lying Rydberg states ($n<15$) the state lifetime is less than $1\,\si{\micro\second}$ meaning that in principle the system is capable of imaging at a million frames per second. In reality the speed at which images can be captured is directly related to the capabilities of the optical camera used. 
If a camera designed for high-speed imaging is used, then it is possible to record THz images at kHz frame rates.
Figure~\ref{fig:chopper} shows frames from a terahertz video taken at 12,000 frames per second. The video shows an optical chopper wheel rotating at 2000 rpm, the white arrow highlights the movement of one spoke of the wheel between frames. This is the fastest demonstration of terahertz imaging to date, 4 times faster than previous demonstrations using this technique~\cite{Downes2020}. 
If such high speeds are not required then most optical cameras are capable of capturing real-time terahertz videos using this technique. 

\subsection{Sensitivity and Minimal Detectable Power}\label{sec:sensitivity}
We characterise the sensitivity of this imaging technique by measuring the minimum detectable power (MDP); 
the minimum terahertz power at which the resulting fluorescence signal is reliably detectable above the noise. 
To do this we record a series of images both with and without the terahertz field for varying terahertz powers.
The response of the vapour will depend on the terahertz intensity, so to generalise we first estimate the power incident per `pixel' in our imaging scheme. 
To do this we assume that the terahertz beam illuminating the atomic vapour is a perfect Gaussian, and that all the lenses are positioned at their focal lengths. We can then use simple Gaussian optics to get an estimate for the size of the beam at the position of the atoms, and hence work out the average intensity at this point. 
We measured the maximum output power of the terahertz source using a VDI Erikson PM5 power meter and used this as an upper bound on the power reaching the atoms. In reality the power incident on the atoms will be lower than this upper bound as there will be losses through the terahertz optical system and reflections from the glass of the vapour cell. 
In the images used the light sheet covers $904\times904$ pixels, which were binned into $4\times4$ superpixels each of size $(40\,\times\,40)$\,\si{\micro\metre}$^2$. 
\begin{figure}[ht]
    \centering
    \includegraphics[width = 0.9\linewidth]{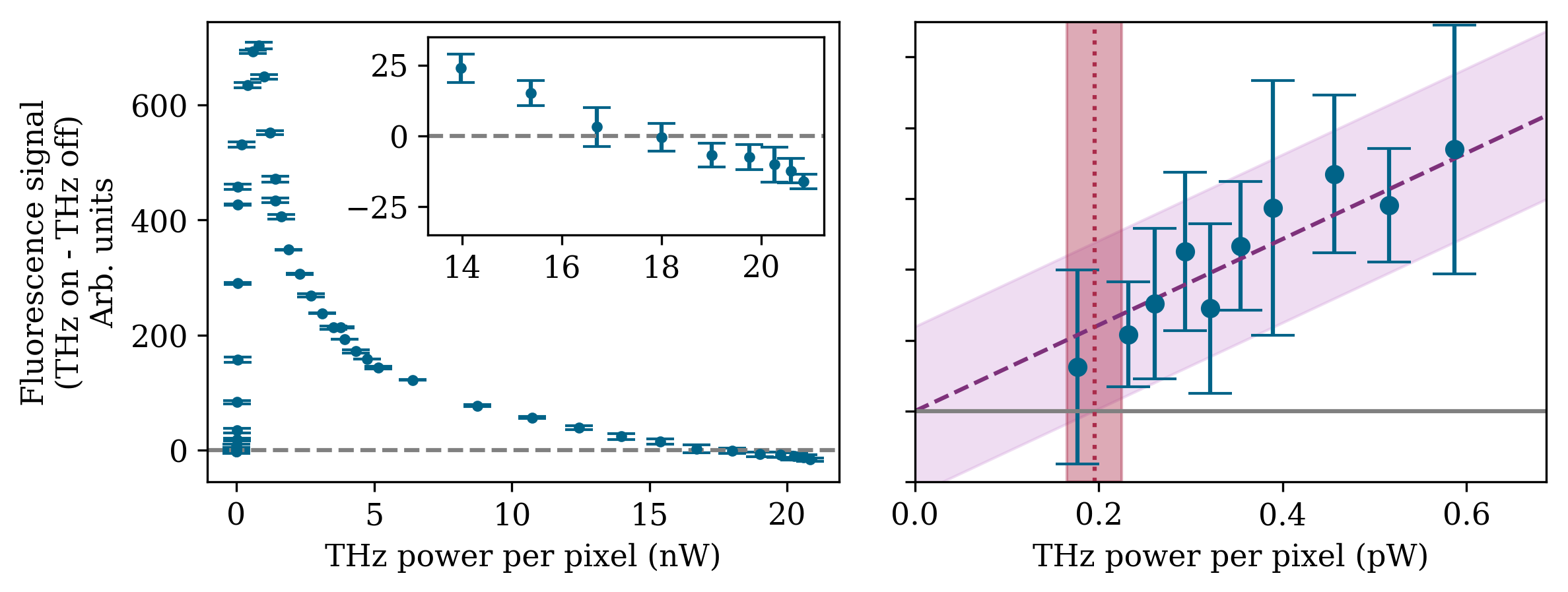}
    \caption{\textbf{Fluorescence Response and Minimum Detectable Power} Left: Background-subtracted counts as a function of applied THz power shows a region of sharp linear increase for low THz powers. This increase stabilises and then counts reduce as THz power increases until eventually the fluorescence in the absence of the THz field is brighter than that in the presence of the THz field (shown in the inset). Right: By fitting to the data the low-power region in which the vapour response is linear we can calculate the MDP as the point at which the signal is no longer reliably detected above the noise. This gives an MDP of $(190\,\pm\,30)$\,\si{\femto\watt\per\second\tothe{1/2}}\, per $(40\times40)$\,\si{\micro\metre}$^2$ pixel for this experimental configuration.}
    \label{fig:MDP}
\end{figure}
We define the `fluorescence signal' as the pixel value resulting from fluorescence in the presence of the terahertz field minus that from background fluorescence in the absence of the terahertz field, $f_{\rm{on}} - f_{\rm{off}}$. 
For a typical superpixel close to the centre of the image, the fluorescence signal and its uncertainty is plotted against the incident terahertz power in Fig.~\ref{fig:MDP}.
Initially at very low terahertz intensities, the fluorescence signal increases linearly with applied terahertz power. 
If the terahertz power is increased further, the fluorescence emitted by the vapour not only plateaus but starts to decrease. This is due to the strong terahertz field causing Autler-Townes splitting of the atomic states \cite{autlertownes1955,anisimov2011}, moving the peak absorption of the excitation lasers away from their resonance position. At very high terahertz powers this can lead to `negative' fluorescence signals being observed, where the amount of fluorescence emitted by the vapour in the presence of the terahertz field is actually less than that in the absence of the terahertz field. This is highlighted in the inset in Fig.~\ref{fig:MDP}. 

We determine the MDP by considering the signal in the region of lowest incident terahertz power where the response is linear, shown on the right hand panel of Fig.~\ref{fig:MDP}. The purple dashed line is a linear fit to the data, with the uncertainty shown by the purple shaded region. 
From this we find an MDP of $(190\,\pm\,30)$\,\si{\femto\watt} per pixel for a 1 second exposure time (red dotted line and shaded region).
At exposure times of over 0.5\,\si{\second} the fluorescence signal saturates the camera, so to obtain an integration time of 1 second we average over 5 frames, each with an exposure of 200\,\si{\milli\second}. 
At this exposure time we are working within the shot-noise limited regime of the camera where the recorded pixel value scales linearly with exposure time, and the uncertainty is proportional to the square root of this pixel value \cite{Hughes:10}. 
In this way the MDP of our system scales inversely with the square root of the total integrated exposure time used, resulting in an MDP of $(190\,\pm\,30)$\,\si{\femto\watt\per\second\tothe{1/2}}\, per $(40\times40)$\,\si{\micro\metre}$^2$ pixel. 
Alternatively this MDP can be expressed as a minimum detectable terahertz intensity of $(0.12\pm0.02)\,\si{\milli\watt\metre}^{-2}\si{\second}^{-1/2}$.

We note again that the numbers quoted here are very much dependent on the experimental configuration used and could be affected by factors such as laser intensity, cell temperature and experiment stability. Different atomic species and excitation pathways may also exhibit different sensitivities and yield different values for the MDP. 

\subsection{Spectral Bandwidth}\label{bandwidth}
  
\begin{figure}[t]
    \centering
    \includegraphics[width = 0.9\linewidth]{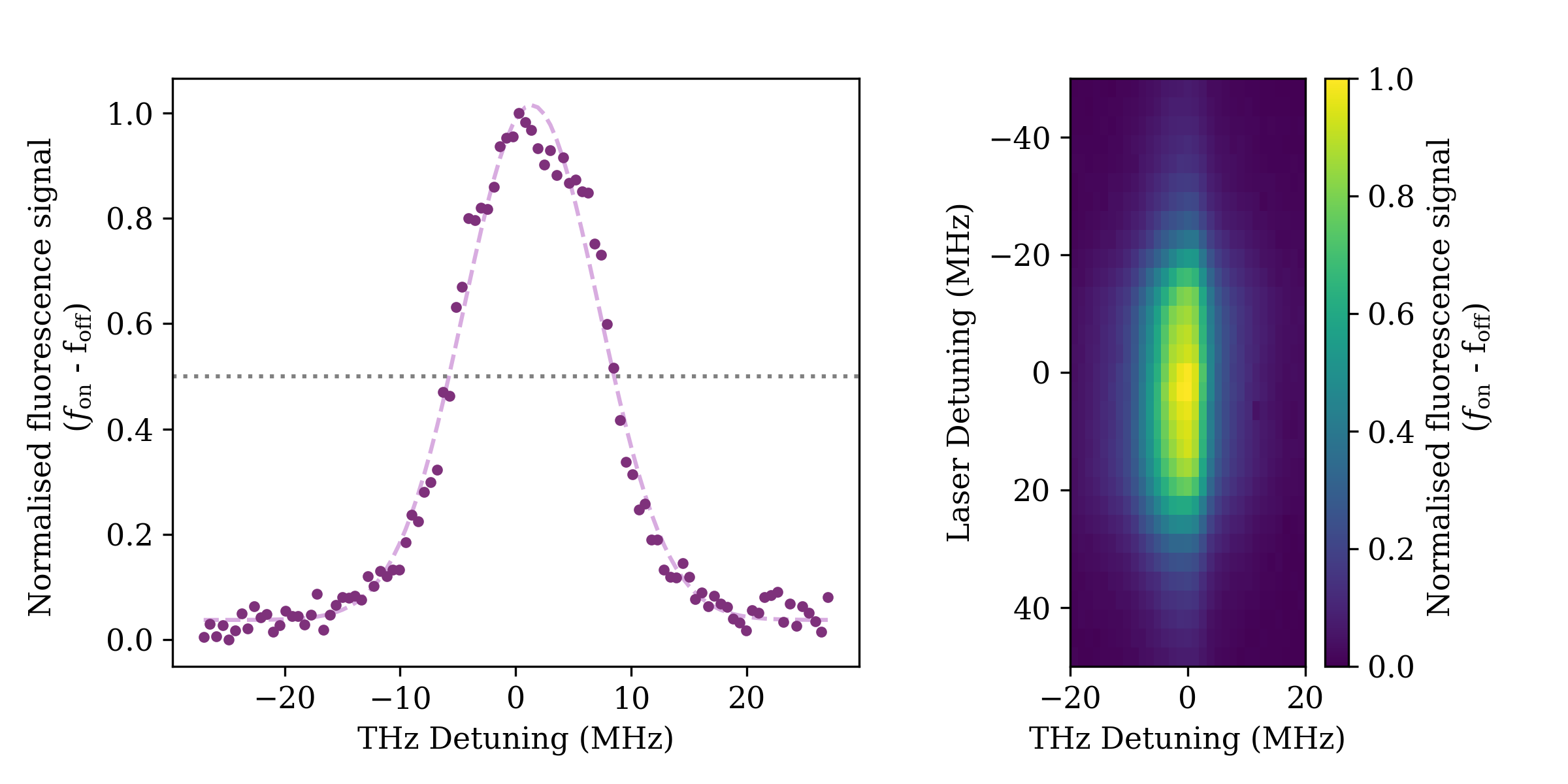}
    \caption{\textbf{Linewidth} Left: Background-subtracted fluorescence signal as a function of THz detuning for the case where all excitation lasers are resonant. Fitting (dashed purple line) reveals a FWHM of $13.8\pm0.2\,\si{\mega\hertz}$. Right: Detuning both the final excitation laser and the THz field reveals the range over which the fluorescence signal is visible. The linewidth of the initial Rydberg transition is wider than that of the THz coupled transition at about 45\,\si{\mega\hertz} at typical experimental conditions.}
    \label{fig:linewidth}
\end{figure}

This terahertz imaging technique requires illumination of an object with a narrowband THz field, but as it relies on atomic transition which has a finite linewidth, there will be a range of terahertz frequencies over which we can still observe a fluorescence signal from the vapour. We can measure the spectral bandwidth of this technique by monitoring the emitted fluorescence as a function of the terahertz frequency. Fig.~\ref{fig:linewidth} shows a typical response, which reveals a full-width half-maximum (FWHM) of $13.8\pm0.2\,\si{\mega\hertz}$. The exact value will differ depending upon several experimental parameters. The main influence on the linewidth is the intensities of the driving lasers; the system is driven above saturation and displays power broadening hence the observed linewidth is greater than the intrinsic atomic linewidth. An applied magnetic field will also influence the observed width and shape of the feature by changing the relative positions of the unresolved hyperfine levels within the Rydberg state \cite{Downes_Thesis}.    
  
The narrow spectral response of the system to applied terahertz radiation further highlights the need for a frequency-stable and tunable terahertz source for this application. The data of Fig.~\ref{fig:linewidth} would be very difficult to obtain using any terahertz source other than an AMC.

\section{Applications and Outlook}\label{sec:applications}

Terahertz imaging is already used for a multitude of applications~\cite{Mittleman18,Roadmap2021}, however up to the point of writing, there is not a terahertz imaging technology available, besides that presented here, capable of imaging at kilohertz frame rates. However, although offering very high speeds and diffraction-limited spatial resolution, atom-based terahertz imaging does not offer the broad spectral information of some other terahertz imaging techniques such as TDS, as it operates CW at a single frequency. Nevertheless, we envisage numerous potential applications for this techniques and, in this section, we provide an outlook for the technology. 

\begin{figure}[hbt]
\begin{centering}
\includegraphics[width=0.9\textwidth]{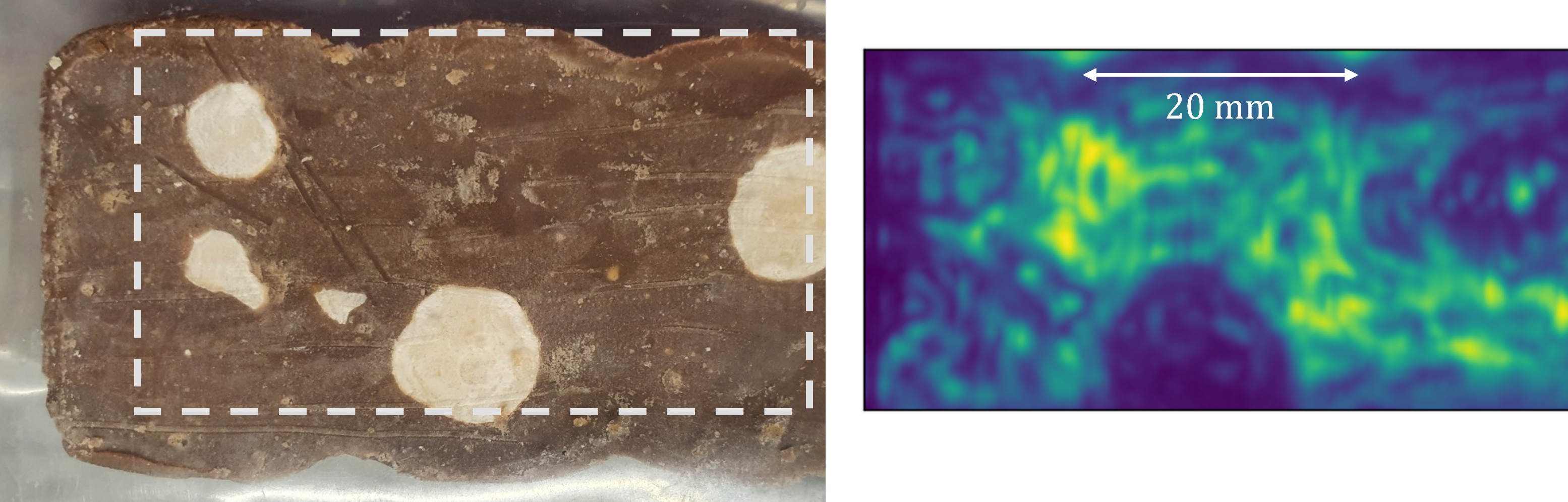}
\caption{\textbf{Food quality inspection.} Optical (left) and terahertz (right) images of inclusions (nuts) within milk chocolate. The nuts and nut fragments appear as darker patches on the terahertz image due to the difference in refractive indices between the nuts and the chocolate. The white dashed square on the optical image indicates the region shown in the terahertz image. Note that in the optical reference image the sample has been sliced to reveal the inclusions. }
\label{fig:choc}
\end{centering}
\end{figure}

The ability to capture an image at high speed, through an optically opaque material, lends itself to applications such as mail security screening and production line monitoring~\cite{Downes19}. In both applications, large numbers of objects pass through a sorting system on conveyors. Terahertz imaging is able to detect non-metallic inclusions or contaminants that other techniques such as x-ray imaging are not able to detect. Figure~\ref{fig:choc} shows an example of terahertz imaging of a chocolate bar with an optical reference image alongside. Terahertz imaging allows for the distribution of inclusions embedded within the chocolate to be examined, in this case hazelnuts.

\begin{figure}[hbt]
\begin{centering}
\includegraphics[width=0.9\textwidth]{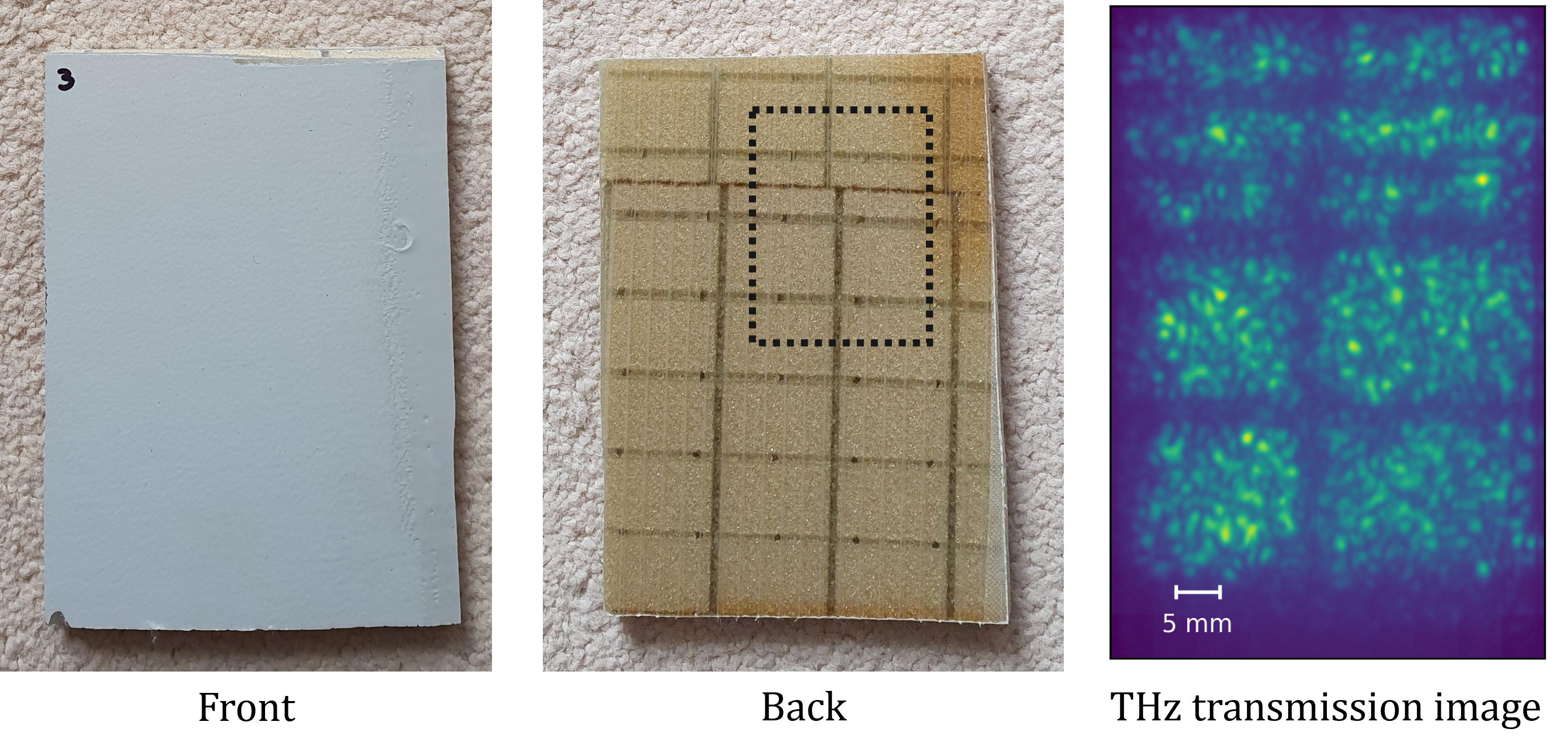}
\caption{\textbf{Imaging of composite materials.} Optical (left and centre) and terahertz transmission (right) images of a segment of wind turbine blade. The dashed black square in the central image indicates the region shown in the terahertz image. The sub-surface structure and discontinuity can be clearly seen in the terahertz image.}
\label{fig:turbine}
\end{centering}
\end{figure}

Other applications may necessitate the inspection of large objects for sub-surface detail. For example, composite materials are widely used in manufacturing and civil engineering. Terahertz waves are able to penetrate deeply into these materials and detect cracks, delamination and other defects below the surface. Figure~\ref{fig:turbine} shows a segment of a wind turbine blade made from composite material (balsa and epoxy). The left and middle panes are optical reference images and the right hand pane is a transmission terahertz image showing sub-surface structure.

All of the images shown here were taken with the imaging system operating in `transmission geometry' where the terahertz field is imaged after passing through an object. It is also possible to use this atom-based imaging scheme to image samples in `reflection geometry', where the image is formed from the terahertz field that is scattered by the surface of an object. Imaging in this modality requires higher terahertz powers as most of the field is absorbed/reflected by the object and so very little power is eventually incident on the vapour cell. 

\begin{figure}[hbt]
\begin{centering}
\includegraphics[width=0.9\textwidth]{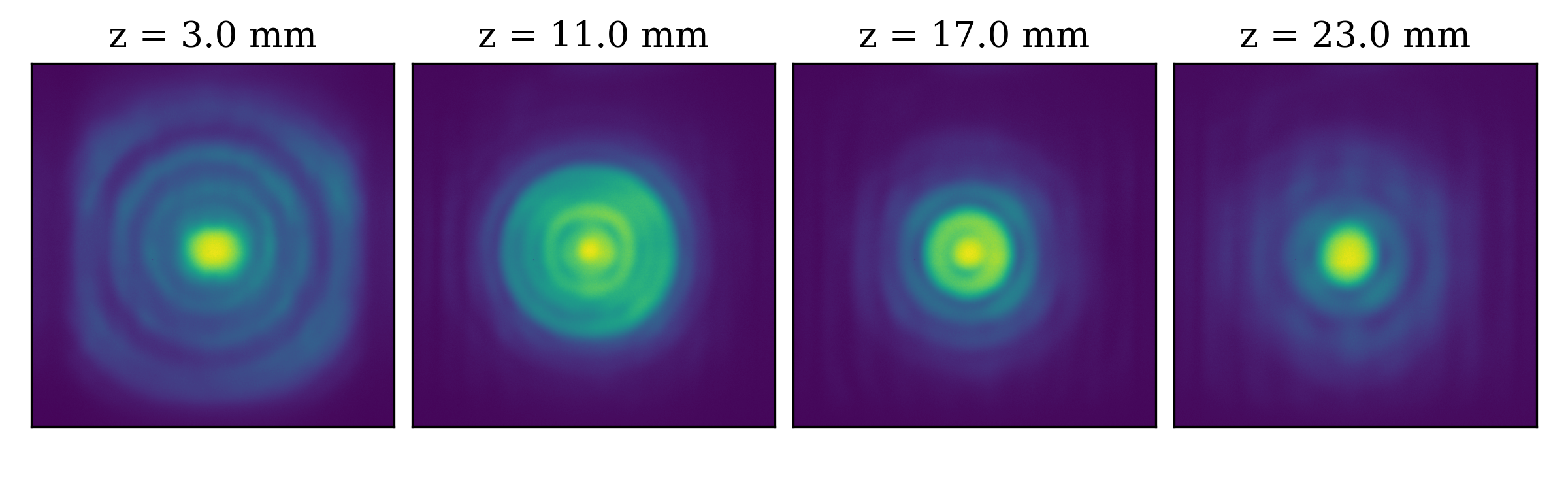}
\caption{\textbf{THz Beam Profiling} Examples of THz beam shape at distance $z$ from the back face of a small (25 mm diameter) hemispherical Teflon lens. The lens clearly shows spherical aberration \cite{F2F}. Vertical fringes arise from interference within the cell and are not due to the lens. Image adapted from \cite{Downes_Thesis}.}
\label{fig:beamprofile}
\end{centering}
\end{figure}

High speed terahertz imaging opens the intriguing prospect of rapid readout and manipulation of the spatial properties of terahertz beams. Terahertz beam profiling is very fast and simple with atom-based imagers. Figure~\ref{fig:beamprofile} show examples of profiling of a terahertz beam after passing through a hemispherical lens. Furthermore, this type of rapid beam profiling allows for real-time adaptive optics (AO) techniques to be applied. Previously AO has been applied to terahertz systems using deformable mirrors~\cite{Brossard:18}, but the low imaging rates allowed only limited utility. Real time wavefront correction could allow for rapid optimisation of terahertz imaging systems where diffraction, interference and distortion are particularly problematic. They could also correct for the poor beam quality of many terahertz sources, particularly QCLs. Recently, we have demonstrated the rapid readout of terahertz orbital angular momentum beams~\cite{Downes2022}, which themselves have potential applications in communications and microscopy. 


We highlight again that terahertz imaging using thermal atomic vapour is a very recent development and there may be many applications, hitherto inconceivable, for terahertz imaging at kilohertz frame rates. The examples discussed above have involved far-field transmission imaging but near-field imaging is also possible with this technique~\cite{Wade2017}. 
Furthermore, the experimental setup presented in Sec.~\ref{sec:exptsetup} largely follows the initial prototype setup developed in Durham. There could be simple gains to be made in the optical engineering and vapour cell design to allow more efficient terahertz-to-optical conversion. There are also ongoing efforts to miniaturise and commercialise the technology, for instance, a compact laser system to perform three-step excitation in caesium for terahertz imaging has been developed with in an academic/industrial collaboration~\cite{Jones:2022}.  
 
\section*{Acknowledgments}

K.J.W. and L.A.D. acknowledge funding from the UK Engineering and Physical Sciences Research Council for grants EP/R002061/1, EP/S015973/1, EP/R000158/1, EP/W033054/1 and EP/T00097X/1 and the European Union's Horizon 2020
research and innovation programme under grant agreement No. 820393.
K.J.W and L.A.D. also acknowledge stimulating discussions with Phil Marsden and Pauline Brown from Unitive Design and Analysis Ltd. as well as Bryn Jones and Adam Selyem from the Fraunhofer Centre for Applied Photonics. We also thank Renewable Advice Ltd. for supplying samples of wind turbine blades and Robert Löw and Danielle Pizzey for help with vapour cell manufacture. 

L.T.C acknowledges the support by Deutsche Forschungsgemeinschaf (DFG) under project grant TO 1142/1-1 and funding from the competence network Quantum-Technology Baden-Württemberg (QTBW.net). Furthermore, L.T.C would like to thank Artur Skljarow and Robert Löw from the University of Stuttgart for their support in the fabrication of a Silicon-glass vapour cell, Xiang L\"{u} and  Holger Grann from the Paul-Drude-Institut für Festkörperelektronik (Berlin) for the design of a QCL for the experiment and for fruitful discussions as well as Eric Dorsch and Felix Battran for their contribution to the research. 

\section*{References}

\bibliographystyle{iopart-num}
\typeout{}
\bibliography{tutorial}

\end{document}